\documentclass[twocolumn]{svjour3}          
\smartqed  
\usepackage{graphicx}
\usepackage{booktabs}
\makeatletter
\expandafter\let\csname opt@amsmath.sty\endcsname\relax
\AtBeginDocument{
	\mathindent=15pt 
	\@mathmargin\@centering} 
\makeatother
\usepackage{amsmath,amsfonts}

\usepackage{color}
\usepackage{multirow}
\usepackage[boxed,ruled,lined,linesnumbered]{algorithm2e}
\usepackage{subfigure}
\usepackage{caption}
\usepackage{tikz}
\usepackage{natbib}
\definecolor{lightseagreen}{rgb}{0.13, 0.7, 0.67}
\usepackage{lipsum}

\newcommand{\pgftextcircled}[1]{
	\setbox0=\hbox{#1}%
	\dimen0\wd0%
	\divide\dimen0 by 2%
	\begin{tikzpicture}[baseline=(a.base)]%
	\useasboundingbox (-\the\dimen0,0pt) rectangle (\the\dimen0,1pt);
	\node[circle,draw,outer sep=0pt,inner sep=0.1ex] (a) {#1};
	\end{tikzpicture}
}

\newcommand{\pgftextcircledblk}[1]{
	\setbox0=\hbox{#1}%
	\dimen0\wd0%
	\divide\dimen0 by 2%
	\begin{tikzpicture}[baseline=(a.base)]%
	\useasboundingbox (-\the\dimen0,0pt) rectangle (\the\dimen0,1pt);
	\node[circle,draw,outer sep=0pt,inner sep=0.1ex,fill=blue] (a) {#1};
	\end{tikzpicture}
}

\definecolor{org}{HTML}{FE8A71}
\definecolor{blu}{HTML}{63ACE5}

\def\Prob{{\bf Prob}}
\def\Proj{{\rm Proj}}

\def\col{{\rm col}}

\def\sign{{\rm sign}}
\def\supp{{\rm supp}}
\def\t0{{t_0}}

\def\E{{\mathbb E}}

\def\R{{\mathbb R}}        
\def\ga{{\gamma}}
\def\la{{\lambda}}

\def\YY{{\mathcal Y}}

\def\grad{{\nabla}}
\def\Prox{{\mathrm{Prox}}}
\def\Proj{{\mathrm{Proj}}}

\newfont{\mycrnotice}{ptmr8t at 7pt}
\newfont{\myconfname}{ptmri8t at 7pt}

\newtheorem{asum}{Assumption}

\newtheorem{defi}{Definition}

\newtheorem{prop}{Proposition}
\newtheorem{Problem}{Problem}
\def\var{{\mathrm{Var}}}
\usepackage[ruled,linesnumbered]{algorithm2e}
\definecolor{lightseagreen}{rgb}{0.13, 0.7, 0.67}
\definecolor{highlight}{HTML}{355C7D}
\usepackage[colorlinks,linkcolor=blue,citecolor=blue]{hyperref}
\newcommand*{\affmark}[1][*]{\textsuperscript{#1}}

 \journalname{International Journal of Computer Vision}
\begin{document}

\title{
Evaluating Visual Properties via Robust HodgeRank
}


\author{Qianqian~Xu\affmark[1]	\and
        Jiechao~Xiong\affmark[2]	\and
        Xiaochun~Cao\affmark[3,4,7]	\and \\
        Qingming~Huang\affmark[1,5,6,7${\star}$]	\and
        Yuan~Yao\affmark[8${\star}$] 
}


\institute{
	Qianqian Xu \at
	\email{xuqianqian@ict.ac.cn}
	\and
	Jiechao Xiong \at
	\email{jchxiong@gmail.com}
	\and
	Xiaochun Cao \at
	\email{caoxiaochun@iie.ac.cn}
	\and
	Qingming Huang \at
	\email{qmhuang@ucas.ac.cn}
	\and
	Yuan Yao \at
	\email{yuany@ust.hk}
	\and
	$^{\star}$Corresponding authors.
}

\date{Received: date / Accepted: date}

\maketitle

\begin{abstract}
 Nowadays, how to effectively evaluate visual properties has become a popular topic for fine-grained visual comprehension. In this paper we study the problem of how to estimate such visual properties from a ranking perspective with the help of the annotators from online crowdsourcing platforms. The main challenges of our task are two-fold. On one hand, the annotations often contain contaminated information, where a small fraction of label flips might ruin the global ranking of the whole dataset. On the other hand, considering the large data capacity, the annotations are often far from being complete. What is worse, there might even exist imbalanced annotations where a small subset of samples are frequently annotated. Facing such challenges, we propose a robust ranking framework based on the principle of Hodge decomposition of imbalanced and incomplete  ranking data. According to the HodgeRank theory, we find that the major source of the contamination comes from the cyclic ranking component of the Hodge decomposition. This leads us to an outlier detection formulation as sparse approximations of the cyclic ranking projection. Taking a step further, it facilitates a novel outlier detection model as Huber's LASSO in robust statistics. Moreover, simple yet scalable algorithms are developed based on Linearized Bregman Iteration to achieve an even less biased estimator. Statistical consistency of outlier detection is established in both cases under nearly the same conditions. Our studies are supported by experiments with both simulated examples and real-world data. The proposed framework provides us a promising tool for robust ranking with large scale crowdsourcing data arising from computer vision.
\keywords{Visual Properties \and Hodge Decomposition \and Linearized Bregman Iteration \and Paired Comparison \and Robust Ranking}
\end{abstract}

\footnotetext[1]{Key Laboratory of Intelligent Information Processing, Institute of Computing Technology, Chinese Academy of Sciences}
\footnotetext[2]{Tencent AI Lab}
\footnotetext[3]{State Key Laboratory of Information Security, Institute of Information Engineering, Chinese Academy of Sciences}
\footnotetext[4]{School of Cyber Security, University of Chinese Academy of Sciences}
\footnotetext[5]{School of Computer Science and Technology, University of Chinese Academy of Sciences}
\footnotetext[6]{Key Laboratory of Big Data Mining and Knowledge Management, University of Chinese Academy of Sciences}
\footnotetext[7]{Peng Cheng Laboratory}
\footnotetext[8]{Department of Mathematics, Hong Kong University of Science and Technology}

\section{Introduction}
\indent  In recent years, we have witnessed a great success toward object category recognition in the computer vision society. However, understanding visual objects often goes beyond  merely knowing their categories. Diving deeper into the visual nature of an object, we can  find fine-grained visual properties about  appearance, shape, color, texture, or some even more complicated contexts. To name a few, understanding human faces requires accurate estimations of visual properties such as age, color properties like \emph{eyes-color}/\emph{face-color}, expression properties like \emph{smile}/\emph{crying}, \textit{etc.}; scene recognition could be regarded as a task to estimate visual properties like \emph{cluttered}/\emph{modern}/\emph{opening area}, \emph{etc.}; predicting user experience of a given image/video is based on the estimation of  visual properties concerning image/video quality. Driven by the urgent demand of fine-grained visual comprehension,
 evaluating such visual properties has become an important building block in many computer vision applications ranging from zero-shot learning \citep{zs1,zs2,zs3}, face editing \citep{faed1,faed2}, age estimation \citep{age1,age2} to image/video quality assessments \citep{tmm12,mm13}.

In this paper, we focus on evaluating visual properties under the primal goal to recommend the top objects with respect to a given visual property from a list of  candidates. Most existing methods along this line  adopt a learning to rank paradigm \citep{ltr1,ltr2}.  More specifically, given a specific visual property and a dataset of object pairs to be evaluated, a ranking function is  learned to  predict the win/lose result of pairwise comparisons in the dataset.


The learning to rank framework brings new possibilities to understanding fine-grained differences in visual properties. However,  the  quadratic dependency on pairwise samples coming with it also makes dataset collection extremely consuming. Thanks to the recent advances of crowdsourcing platforms (e.g.,
\href{https://www.mturk.com}{MTurk},
\href{http://www.innocentive.com/}{Innocentive},
\href{http://crowdflower.com/}{CrowdFlower},
\href{http://www.crowdrank.net/}{CrowdRank}, and
\href{http://www.allourideas.org/}{Allourideas}),
collecting pairwise dataset becomes much  easier with the collaborative efforts of the online  annotators \citep{li2016crowdsourced}. More precisely, the object pairs to be compared are simultaneously assigned to a large group of online  annotators. The annotation process could then be carried out much more efficiently in such a distributed fashion.  However, learning pairwise rankings in crowdsourcing scenario also brings new difficulties inherently: (i) incomplete and imbalanced data  source:  collecting a complete set of comparison labels is almost impossible, worse one might find some of the pairs are frequently annotated while the others seldomly are clicked by anyone; (ii) conflicts of interests or inconsistencies in the data, where different annotators might provide different results regarding a same comparison, due to either subjective factors or quality factors.


\begin{figure}
	\begin{center}
		\includegraphics[width=1 \columnwidth]{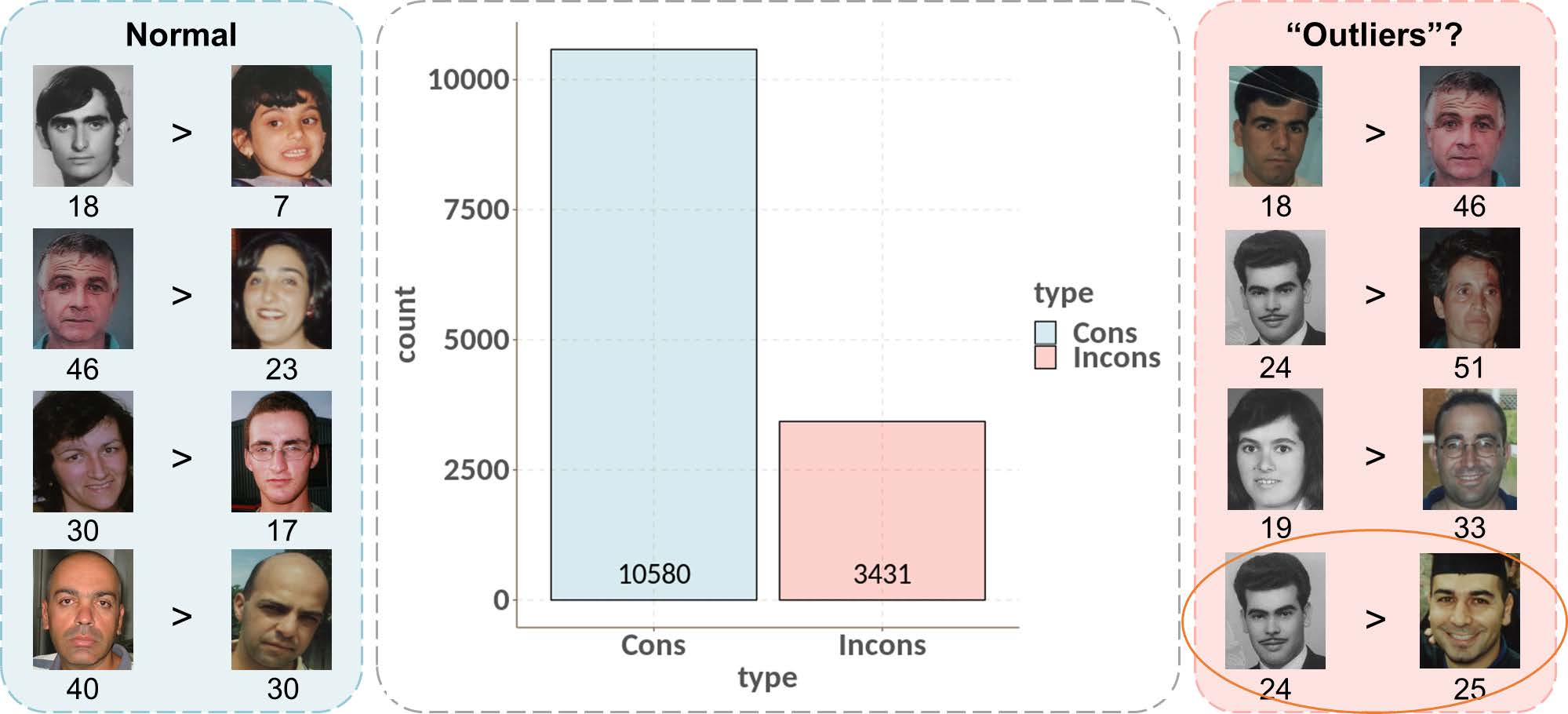}
		\caption{Annotation Distribution on Human Age Dataset: As a surprising fact, we find that 3431 out of 14011 annotations (account for 24.5\% comparisons in the total dataset) are actually inconsistent with the ground-truth. Note that since there is no guarantee of the annotation quality, trivial outliers such as the first three examples shown here are by  no means rare. This motivates us to define outliers in a data-driven way instead of simply using the semantic hardness of ranking. }\label{fig:stat}
	\end{center}
\end{figure}

During the past decades, a great number of studies \citep{Hodge,MM11,tmm12,Kahle09,Kahle14,MM12,TMM13,RM1, RM2} have been successfully carried out to explore learning to rank problem in the context of crowdsourcing. However, there is an overlooked issue about the untrustworthy data in the crowdsourcing platforms.  In fact, due to the lack of supervision in crowdsourcing experiments, there is no guarantee for the annotation quality \citep{MM09}. For example, when the testing time for a single participant lasts too long, the participant may become impatient and  input random decisions. Such random decisions are typically inconsistent with majority decisions from the crowd. In fact, such inconsistent results are ubiquitous in crowdsourced datasets. To see this, we  illustrate our statistical observations on a specific dataset concerning human age estimation. In this dataset, we asked annotators to annotate whether a given human $x$ is older than another human $y$.  As shown in Fig.\ref{fig:stat}, it turns out that 3431 out of 14011  annotations (account for \emph{24.5\%} comparisons in the total dataset) are actually inconsistent with the ground-truth. The reason behind this subtle fact is that the annotation quality is constantly affected by: 1) impatient users who usually provide wrong results due to carelessness; 2) malicious users who provide adversarial results; 3) zombie accounts which could only give random results; 4) the multi-criteria nature of human decision that different people tend to adopt different criteria (color/wrinkle/beard/hair) when making their own decisions. Since inconsistent results are ubiquitous and complicated, a crucial problem then comes as \emph{what are the real outliers in the dataset that may significantly affect statistical aggregations of ranking data and need a different treatment than stochastic noise}. In a naive way, outliers can be defined as all the annotations that have an unusual deviation (inconsistency) from the most common or expected pattern. Unfortunately, we will soon see from our example that this is not a proper choice. In the right side of Fig.\ref{fig:stat}, we show four typical instances of inconsistent annotations. Generally speaking, they include two kinds of results: 1) in the first three annotations, the results are flipped when the age difference is significant; 2) the last instance in circle gives inconsistent annotation when such difference is not too easy to be recognized. Annotations of the first kind significantly influence the global ranking and thus should be recognized as outliers. By contrast, inconsistency in the second kind of results is mainly due to the fact that the underlying pair of objects are, in themselves, too similar to distinguish, that can be regarded as stochastic noises. Consequently, the definition of outliers should exclude unavoidable statistical errors. In most applications, the ground truth is unknown (unlike the example above) and the \emph{identifiability} of such outliers thus becomes a challenge. In this paper, our goal is then to seek out proper models that could learn a global consistent ranking with such outliers identified effectively and efficiently.

In this paper, we particularly bring two sets of tools with different algorithms, one from robust regression as Huber's LASSO, the other from computational mathematics known as Linearized Bregman Iteration (LBI). Our foundation is based on the Huber's robust regression \citep{Huber81}, which combines the least squares for stochastic noise and the least absolute deviation for outliers. Such a robust regression is equivalent to a LASSO problem \citep{LASSO_new1,LASSO_new2}, which enables us to pursue outlier identification based on modern variable selection techniques, e.g., \citep{Owen}, with applications in crowdsourced QoE assessment of multimedia \citep{mm13}. However, it is well-known that LASSO is biased \citep{Fan2001}. Moreover, since the amount of outliers is typically unknown, it is desired to solve the full LASSO regularization path which is too expensive in crowdsourcing applications with thousands of outliers or more. This drives us to the second model which is a  simple iterative algorithm based on LBI. For outlier detection, we provide theoretically guarantees the outlier identifiability for both Huber's LASSO and LBI. Beyond such a theoretical attraction, LBI generates the whole sparse regularization path along its simple iterations, which combines an iterative gradient descent algorithm together with a soft thresholding, at nearly the same cost of computing a LASSO estimator at a fixed regularization parameter using the well-known Iterative Soft-Thresholding Algorithm (ISTA) (e.g., \citep{fista,fista2,fista3} and references therein) which converges to the biased LASSO solution. To obtain a regularization path, ISTA needs to run with many different thresholding parameters till convergence to follow the biased LASSO path; in contrast, our LBI algorithm only needs a single run and regularization is achieved by early stopping like boosting-type algorithms, which may save the computational cost greatly and thus suitable for large scale implementation, e.g., distributive computation.


As a summary, our main contributions are highlighted as follows:
\begin{itemize}
	\item[(A)] A robust HodgeRank framework is proposed to deal with outliers in the crowdsourcing platforms
	with outlier identification as sparse approximations of the cyclic ranking projection in Hodge decomposition.
	\item[(B)] We also show that the sparse outliers can be identified by a LASSO problem from robust statistics, i.e., Huber-LASSO, where its regularization paths provide us an order
	on samples tending to be outliers.
	\item[(C)] Scalable algorithms are proposed based on Linearized Bregman Iteration which is suitable for large scale analysis and may render less biased estimates.
	\item[(D)] Identifiability of outlier is established for both Huber-LASSO and Linearized Bregman Iteration as statistical model-selection consistency under nearly the same set of conditions.
\end{itemize}

This paper is an extension of our conference paper \citep{mm13}, which formulates the outlier detection as a LASSO problem. However, the computational cost of full LASSO path is prohibitive from the applications of large scale crowdsourcing problem. As a departure from the previous work, here we consider the Linearized Bregman Iteration for outlier detection, which is motivated by some dynamics (or more precisely, inclusion) to remove bias in LASSO paths and allows extremely simple and scalable implementations.

The remainder of this paper is organized as follows. In Section \ref{sec:re}, we provide a survey of related work.  We then systematically introduce the principle of robust HodgeRank and algorithms, based on the Huber-LASSO and Linearized Bregman Iterations in Section \ref{sec:Outlier_Detection}. Statistical consistency theory for both algorithms is established in Section \ref{sec:consistency}. Detailed experiments are presented in Section
\ref{sec:experiments}, followed by the conclusions in Section \ref{sec:conclusions}.

\section{Related work}\label{sec:re}
\subsection{Crowdsourced Ranking Methods}
Historically in crowdsourcing study \citep{MM09,Wu13crowd} of multimedia,
Transitivity Satisfaction Rate (TSR) serves as a typical method for outlier detection, which checks all the
intransitive (circular) triangles such that $A\succ B \succ C\succ A$ where $A\succ B$ means $A$ has a higher rank than $B$. The TSR is defined as the
number of transitive judgment triplets divided by the total
number of triplets where transitivity may apply. The drawback of TSR  is that it can only be applied for complete and balanced paired comparison data. When the paired data are incomplete, i.e., have
missing edges, it remains open how to detect the noisy pairs.

Fortunately, a recent Hodge-theoretic approach to statistical ranking \citep{Hodge} called HodgeRank, comes out as a solution with its ability to characterize the intrinsic inconsistency in pairwise data geometrically while inferring a global ranking from incomplete and imbalanced samples. Specifically, it solves the problem with an orthogonal decomposition of incomplete and imbalanced pairwise ranking data into global ranking and cyclic rankings that consists of both globally cyclic (harmonic) and triangular cyclic rankings. Cyclic rankings are due to either stochastic noise or outliers that have large deviations from the population and consequently of high influence on the estimation of global rankings. Such insight enables us to treat the outlier identification as \emph{sparse approximations of cyclic ranking projection in Hodge decomposition}.

Practically, such a methodology has been rapidly applied to the crowdsourced assessment for quality of experience (QoE) in multimedia, together with random graph models of sampling design \citep{MM11,tmm12}, which is a favorable choice for crowdsourcing studies where individuals provide the ratings in more diverse and less controlled settings than traditional lab environment \citep{added}. With recent developments on topological random graph theory in mathematics \citep{Kahle09,Kahle14}, the framework of HodgeRank on Random Graphs enables us to derive the constraints on sampling complexity to which the random selection must adhere. Instead of pursuing a quadratic number of pairs ${n \choose 2}\approx n^2$ in a complete paired comparison design for $n$ candidates, Erd\"{o}s-R\'{e}nyi random graph designs only need almost linear $\Omega(n\log n)$ distinct sample pairs to ensure the graph connectivity for inferring a global ranking score and $\Omega(n^{3/2})$ distinct random edges are sufficient to
remove the global inconsistency. Moreover, it allows an extension to online sampling settings. In \citep{MM12,TMM13}, some online algorithms are proposed based on the classic Robbins-Monro procedure \citep{RM1, RM2} or stochastic gradient method for HodgeRank on random graphs, together with online tracking of topological evolution of paired comparison (clique) complex. For edge independent sampling processes, the online algorithm for global rating scores reaches the minimax optimal convergence rate hence asymptotically as efficient as a batch algorithm, when dealing with streaming or big data. Most recently \cite{outxu} propose $\ell_0$-norm-based sparse approximation on top of HodgeRank to deal with the outlier patterns existed in the pairwise crowdsourcing data.

However, except the recent work in  \cite{outxu}, the majority of the related work ignore the outliers existed in the crowdsourcing systems. As shown in the introduction, the outliers are by no means rare nor do they easy to be detected. Defining and finding out outliers then become our major task in this paper. Compared with the recent work \cite{outxu} which also deals with outliers, our proposed algorithm is both sign-consistent and bias-free. Moreover, our LBI method also provides us a new way for model selection which simultaneously learns the model parameters and chooses the hyper-parameters within a single run of a simple iterative algorithm.

\subsection{Linearized Bregman Iteration}

In this paper we borrow the strength of  a powerful algorithm known as Linearized Bregman Iterations (LBI). LBI was firstly introduced in \citep{YODG08} in the literature of applied mathematics, as a simplified scheme for Bregman iteration \citep{OBG+05} which may reduce bias, also known as contrast loss, in the context of Total Variation image denoising. Recently with the aid of its limit dynamics, a system of differential inclusions called \emph{Inverse Scale Space}, it is shown \citep{osher2014} that 
for sparse linear regression such algorithms may reach model selection consistency under nearly the same conditions as LASSO, yet produce the \emph{unbiased Oracle} estimator or an arbitrary good approximation. Therefore, \citep{osher2014} steps further to study  noisy sparse signal recovery via a differential equation approach induced by LBI which may return an estimator that is both sign-consistent and bias-free. Most recently, the theoretical strength of LBI has been further analyzed in \cite{lbiglobal,dynamic} and has been applied to few-shot and zero-shot learning \cite{zs2}.

Different with the existing work, we provide a specific application of the LBI algorithm to the field of outlier robust crowdsourced ranking. Beyond this, we also provide a specific definition of the outliers in this context based on the Hodge theory. Moreover, our new theoretical results in Thm.\ref{thm:LASSO} and Thm.\ref{thm:LB} are specifically designed for our task.

\section{Robust HodgeRank} \label{sec:Outlier_Detection}

In this section, we systematically introduce the Hodge Decomposition of pairwise ranking data, called HodgeRank here, and the Robust HodgeRank with outlier detection as sparse approximation of cyclic rankings. Various techniques from robust statistics  can be used here (e.g., Huber-LASSO). Finally we introduce some scalable algorithms using linearized Bregman iterations.

Given a specific visual property, let ${\wedge}$ be a set of raters and ${V=\{1,...,n\}}$ be the set of items to be ranked. Paired comparison data is collected as a real-valued function with missing values, $\wedge\times V \times V\to \R$, which is \emph{skew-symmetric} for each $\alpha\in \wedge$, i.e., $Y^\alpha_{ij}=-Y^\alpha_{ji}$ representing the degree that rater $\alpha$ prefers $i$ to $j$. Without loss of generality, one assumes that $Y_{ij}^\alpha>0$ if $\alpha$ prefers $i$ to $j$ and $Y_{ij}^{\alpha}\leq 0$ otherwise. The choice of scale for $Y_{ij}^\alpha$ varies in applications. For example, in multimedia quality of experience assessment  \citep{MM11}, dichotomous choice $\{\pm 1\}$ or a $k$-point Likert scale ($k=3,4,5$) are often used; while in information retrieval \citep{CorMohRas07} and surface reconstruction \citep{Yu12} etc., general real values are assumed.

Such pairwise comparison data can be represented by a graph $G=(V,E)$, where $(i,j)\in E$ is an oriented edge when $i$ and $j$ are effectively compared by some voters. Associate each $(i,j)\in E$
an Euclidean space $\R^{|\wedge_{ij}|}$ where $\wedge_{ij}$ denotes the voters who compared $i$ and $j$. Now define $\YY:=\otimes_{(i,j)\in E} \R^{|\wedge_{ij}|}$, 
an Euclidean space with standard basis $e_{ij}^\alpha$. It can be equivalently viewed as a multigraph with directed edges.

\subsection{Hodge Decomposition and Sparse Cyclic Approximation}
Now we introduce Hodge Decomposition for pairwise ranking on graphs \citep{Hodge}. Roughly speaking, it says that all paired comparison data $\YY$ on graph $G$ admits the following orthogonal decomposition:
\begin{equation}\label{eq:proj}
\YY = global\ ranking \oplus cycles,
\end{equation}
where the component \emph{cycles} can be further decomposed into
\[cycles = local\ cycles \oplus global\ cycles. \]
Here local cycles are bi-cycles ($i\succ j$ and $j\succ i$) and triangular cycles, e.g., $i \succ j \succ k\succ i $; while global cycles are loops involving nodes more than three (e.g., $i \succ j \succ k \succ...\succ i$) and typically traversing all nodes in the graph. These cycles may arise due to conflicts of interests in ranking data. Details are to be shown soon.

In a crowdsourcing task, we assume that most of the workers will vote rationally according to his/her common preference or utility, while occasionally irrational behaviors might happen due to the perturbation in environmental or psychological conditions. For example, the worker may simply click one side or randomly during a period of the test. Such outliers, which can be encoded as a sparse vector $\gamma \in \YY$ consisting a small set of nonzero elements, however may contribute a significant part of the cycles leading to the conflicts of interests. Disclosing such outliers will be our main concern in this paper. With the aid of Hodge decomposition, suppose $\Proj_{cycle}$ is the projection onto the cyclic ranking subspace, then
\begin{equation} \label{eq:outlier}
\Proj_{cycle}(y)=\Proj_{cycle}(\gamma)+\varepsilon
\end{equation}
where $\varepsilon$ is a sub-Gaussian type noise. Additional structures on $\gamma$ might be imposed, such as being user and location specific \citep{ICML16}. From simplicity, in this paper we mainly investigate the scenario of sparse $\gamma$, most of whose elements are zeros. This leads to an extension of outlier detection by TSR in a complete case to incomplete and imbalanced samplings.

\subsection{Problem Definition}
With the HodgeRank framework, we now provide a formal definition of the outliers and then pose the research problem in this paper.
Recall the orthogonal decomposition of the ranking score $\mathcal{Y}$ in Eq.(\ref{eq:proj}).
The cycle ranking subspace ${cycles}$  contains the unexpected outliers that contribute to a cyclic ranking, either in a local form (bicycles/triangles)  $i \succ j \succ i$ / $i \succ j \succ k \succ i$  or in a global form which involves more than three objects  $x \succ j \succ k \succ l \succ \cdots \succ i$. This leads to the following definition of the outliers.
\begin{defi}
\emph{Outlier refers to  the comparisons that facilitate either local or global cycles}. From a practical point-of-view, it consists of comparison annotations that are inconsistent with the majority opinions from the data.
\end{defi}

With the definition of outliers, the robust ranking problem then becomes clear.
\begin{Problem}[Robust HodgeRank]\label{prob}
	Given an input matrix  pairwise annotations $\{y^\alpha_{ij}\}_{\alpha,i,j}$, where $y^\alpha_{ij}$ is the  annotation from user $\alpha$ for a given pair of items $(i,j)$, can we find a proper algorithm to {\color{blue} 1)} learn a global ranking score in $\mathcal{Y}_{global}$ for each item  and {\color{blue} 2)} learn an outliers indicator parameter such that Eq.(\ref{eq:outlier}) holds approximately?
\end{Problem}

\indent In the next two subsections, we will develop two algorithms to solve the given problem:
\begin{enumerate}
	\item[(1)] \textbf{Huber-LASSO}: First of all, we propose a novel algorithm called Huber LASSO, which gives a partial penalized LASSO formulation of the Huber robust regression model. It is interesting to note that the Huber LASSO realizes Eq.(\ref{eq:proj}) via a sparse approximation of $\Proj_{cycles}(\mathcal{Y})$ .
	\item[(2)] \textbf{LBI}: Though HLASSO provides a good approximation of the outliers, it could not choose the model hyperparameters by itself. At the first glance, the cross-validation scheme should be a good candidate for hyperparameter selection.  However, we find it might fail when outliers become dense and small in magnitudes. This motivates us to propose another algorithm LBI, which iteratively generates a sparse regularization path for HLASSO problem. Along the path, LBI could choose a proper  hyperparameter automatically and thus make the hyperparameters of HLASSO tunable. As another point, we also show that the limiting dynamics of LBI could give unbiased estimation of the model parameters, which could not be realized by traditional LASSO formulation.
\end{enumerate}

\subsection{Huber-LASSO}
To deal with sparse outliers, Huber \citep{Huber73,Huber81} proposes the following robust regression,
\begin{equation} \label{eq:huber}
\min_{\sum_{i\in V} \theta_i=0} \sum_{i,j,\alpha}  \rho( \theta_i -  \theta_j - Y_{ij}^\alpha),
\end{equation}
where $\rho:\R\to[0,\infty)$ is a differentiable convex function with derivative $\psi(x)=\rho^\prime(x)$. For fixed number of items $n$, as the number of paired comparisons $m$ satisfies $n/m\to 0$, a solution of (\ref{eq:huber}) satisfies asymptotically $\hat{\theta}\sim N(0, \Sigma)$ where $\Sigma=V(\psi,F_z) L_0^\dagger$ where $V(\psi,F_z) = \E[\psi^2]/(\E[\psi^\prime])^2$. Had we known the density of noise $d F_z (x)= f_z(x) dx$, the optimal choice of $\rho$ would be decided by $\psi(x) = (\log f_z(x))^\prime$ such that the asymptotic variance meets the lower bound via the Fisher information, i.e., $V(\psi,F_z)=1/I(F_z)$. Square loss is hence an optimal choice for Gaussian-type noise. 
When sparse outliers exists, the Huber's loss function $\rho_\lambda$ is an alternative choice for $\rho$
\begin{equation*}
\rho_\lambda(x) =
\left\{
\displaystyle \begin{array}{ll}
x^2/2, & \textrm{if $|x|\leq \lambda$}\\
\lambda |x| - \lambda^2/2, & \textrm{if $|x|> \lambda$.}
\end{array}
\right.
\end{equation*}
It is a strongly convex and smooth function.
When $|\theta_i -  \theta_j - Y_{ij}^\alpha| < \lambda$, the comparison is regarded as a ``good" one with Gaussian noise and $L_2$-norm penalty is used on the residual. Otherwise, it is regarded as a ``bad'' one contaminated by outliers and $L_1$-norm penalty is applied which is less sensitive to the amount of deviation. So when $\lambda\to0$, it reduces to a least absolute deviation (LAD) problem or $L_1$-norm ranking \citep{osting2013statistical}.

Though Huber robust regression has reasonable statistical properties, it is hard to gain intuition on how it helps to remove out outliers in the original form.  Fortunately, one can show that Huber's robust regression \eqref{eq:huber} with $\rho_\lambda$ is equivalent to the following LASSO formulation \citep{mm13}:
\begin{eqnarray} \label{eq:HLasso}
&\min_{\sum_{i\in V}  \theta_i=0,\gamma}\frac{1}{2} \|Y - X \theta - \gamma \|_{2}^2 + \lambda \| \gamma \|_{1} \\
&:= \sum_{i,j,\alpha} [\frac{1}{2}(\theta_i - \theta_j + \gamma_{ij}^\alpha - Y_{ij}^\alpha )^2 + \lambda | \gamma_{ij}^\alpha |]\nonumber,
\end{eqnarray}
which, as we will soon see, provides an explicit way to filter out outliers by virtues of the sparse parameter $\gamma$.
To see the equivalence, let $(\hat{\theta}^{lasso}, \hat{\gamma}^{lasso})$ be a solution of \eqref{eq:HLasso}. Here we introduce a new variable $\gamma_{ij}^\alpha$ for each comparison $Y_{ij}^\alpha$ such that ${|\gamma_{ij}^\alpha|>0}$ is equivalent to $|\hat{\theta}^{lasso}_{i} -\hat{\theta}^{lasso}_{j} -Y_{ij}^\alpha|>\lambda$, i.e., an outlier. To be less sensitive to outliers, an $L_1$-norm penalty of $\gamma_{ij}^\alpha=\hat{\theta}^{lasso}_{i} -\hat{\theta}^{lasso}_{j} -Y_{ij}^\alpha$ is used as in Huber's loss \citep{Huber81}. Otherwise, an $L_2$-norm is used to attenuate the Gaussian noise. This optimization problem is a partially penalized LASSO \citep{lasso}, hence called Huber-LASSO (or HLASSO) in this paper.

HLASSO can be further split into two subproblems with the two groups of variables decoupled, via orthogonal projections of data $Y$ onto the column space of $X$ and its complement. In particular, the outlier $\gamma$ is involved in a standard LASSO problem, whose design matrix comes from the projection onto the complement of the column space of $X$. Some LASSO packages, such as {\tt glmnet 2.0-5}, can directly solve HLASSO \eqref{eq:HLasso}; while other packages such as {\tt quadrupen} cannot deal with partially penalized LASSO, hence one may turn to the standard LASSO sub-problem for $\gamma$.

Precisely, let $X$ has a full SVD decomposition
\begin{equation}
X =  U \Sigma V^T = [U_1, U_2]\begin{bmatrix}
\Sigma_r,& 0 \\
0,& 0 \\
\end{bmatrix}V^\top.
\end{equation}
Here, $U_1$ is an orthonormal basis of the column space $\col(X)$ and  $U_2$ forms an orthonormal basis  for the orthogonal complement  $Ker(X^\top)$. Since $\col(X)$ corresponds to the graph gradient operator, i.e., $\col(X) = global \ ranking$, we have  $Ker(X^\top) = cycles$. In this way, the projector onto $global \ ranking$ and $cycles$ could be represented as:
\begin{align}
\Proj_{global}(\cdot) = U_1U_1^\top(\cdot), \  \Proj_{cycles}(\cdot) = U_2U_2^\top(\cdot)
\end{align}
respectively. Through simple algebraic operations, the above argument yields  the following proposition.
\begin{prop}\label{prop:hlasso}
	The HLASSO solution $(\hat{\theta},\hat{\gamma})$ can be obtained by solving the following two problems sequentially:
	\begin{equation} \label{eq:HLASSO22}
	\min_{\gamma} \frac{1}{2} \| U_2^T  Y - U_2^T \gamma \|_2^2 + \lambda \|\gamma\|_{1}
	\end{equation}
	\begin{equation} \label{eq:HLASSO21}
	\min_{\sum_{i\in V}\theta_i=0} \frac{1}{2} \| U_1^T X \theta - U_1^T(Y-\hat{\gamma}) \|_2^2.
	\end{equation}
	Moreover (\ref{eq:HLASSO22}) and (\ref{eq:HLASSO21}) are equivalent to
	\begin{equation} \label{eq:Heq1}
	\min_{\gamma} \frac{1}{2} \| \Proj_{cycles}(Y) - \Proj_{cycles}(\gamma) \|_2^2 + \lambda \|\gamma\|_{1}
	\end{equation}
	and
	\begin{equation} \label{eq:Heq2}
	\min_{\sum_{i\in V}\theta_i=0} \frac{1}{2} \|  X \theta - \Proj_{global}(Y-\hat{\gamma}) \|_2^2,
	\end{equation}
	respectively.
\end{prop}

{{Note that since
\begin{equation} \label{eq:projection}
U_2U_2^T = I - U_1U_1^T =  I-X(X^TX)^{\dagger}X^T,
\end{equation}
where $A^\dagger$ denotes the Moore-Penrose inverse of $A$, we do not need the full SVD to solve \eqref{eq:HLASSO22}.}}

According to Proposition \ref{prop:hlasso}, the original HLASSO could be  splited into two separate optimization problems: the first problem \eqref{eq:HLASSO22} is a standard LASSO problem which detects outliers in the cycles ranking subspace while the second one \eqref{eq:HLASSO21} is a least squares problem for the corrected data $\Proj_{global}(Y-\hat{\gamma})$ in the global ranking subspace.

According to  (\ref{eq:Heq2}),  HLASSO realizes Problem \ref{prob}-{\color{blue} 1)} by seeking a proper score $\theta$ specifically  on the global ranking subspace. Moreover, in view of  (\ref{eq:Heq1}), HLASSO realizes Problem \ref{prob}-{\color{blue} 2)}  via LASSO: the outlier $\gamma$ in (\ref{eq:HLASSO22}) is a \emph{sparse approximation of the projection of paired comparison data onto cyclic ranking subspace}. In this way, the proposed HLASSO realizes Problem 1 with theoretical guarantees.\\
\indent As what is stated in the introduction, outliers should not include statistical errors.  It is then interesting to note that the sparsity of $\gamma$ realizes this goal.  Since statistical errors typically have a small component on the cycle ranking subspace, then the magnitude of $\gamma$ is likely to dominate the objective function. In this case, the corresponding entry of $\gamma$ will probably be set as 0 so as to achieve the global minimum. Hence the annotations having only statistical errors will not be included in the support set of $\gamma$ with high probability.

\subsubsection{Parameter Tuning}
A crucial question here is how to choose $\lambda$ which is equivalent to estimate the variance of $\varepsilon_{ij}^\alpha$ properly.
For this purpose, Huber \citep{Huber81} proposes the concomitant scale estimation, which jointly estimates $s$ and $\lambda$ via the following way:
\begin{equation}\label{eq:huber2}
\min_{\sum_{i\in V} \theta_i=0, \sigma} \sum_{i,j,\alpha}  \rho_M\left(\frac{\theta_i - \theta_j - Y_{ij}^\alpha}{\sigma}\right)\sigma+m\sigma,
\end{equation}
where $m$ is the total number of paired comparisons. Note that $M\sigma$ plays the same role of $\lambda$ in (\ref{eq:huber}), since for fixed $\sigma$, minimization problem (\ref{eq:huber2}) is equivalent to minimize (\ref{eq:huber}) with $\lambda=M\sigma$. In practice, \citep{Huber81} suggests to fix $M=1.35$ in order to be robust as much as possible and do not lose much statistical efficiency for normally distributed data. Problem (\ref{eq:huber2}) becomes a convex optimization problem jointly in $\theta$ and $\sigma$, hence can be solved efficiently. In LASSO's formulation, Huber's concomitant scale estimation becomes scaled-lasso whose consistency is proved in \citep{SunZha12}.

However, in our applications the concomitant scale estimation (\ref{eq:huber2}) only works when outliers are sparse enough. Moreover, cross-validation to find optimal $\lambda$, turns out to be highly unstable here. Since every sample is associated with an outlier variable, leaving out samples thus loses all information about the associated outlier variables.

Here we suggest a new cross-validation scheme based on random projections. Note that $U_2$ is a projection onto the subspace $\ker(X^T)$, hence one can exploit subsets of random projections as training and validation sets, respectively. Each random projection will contain information of all the sample and outliers generically. Thanks to the exploitation of the well-known Erd\"{o}s-R\'{e}nyi random graphs in crowdsoucring experiments \citep{MM11,tmm12}, positions of outliers can be consistently identified with such a random projection based cross-validation.

In practice, although cross-validation works well for sparse and large enough outliers, we find it might fail when outliers become dense and small in magnitudes. However, when cross-validation fails, we still find it more informative by looking at the regularization paths of (\ref{eq:HLASSO22}) directly. The order that variables $\gamma_{ij}^\alpha$ become nonzero when regularization parameter $\lambda$ changes from $\infty$ to small, can faithfully identify the tendency that a measurement $Y_{ij}^\alpha$ is contaminated by outliers, even when cross-validation fails. Therefore, we suggest to use regularization paths to inspect the outliers in applications.

Prior knowledge can also be used to tune the regularization parameter. For example, if one would like to drop a certain percentage of outliers, say $5\%$, then the top $5\%$ variables appeared on regularization paths can be regarded as outliers and dropped. Moreover, the deviation magnitudes sometimes can be used to determine outliers. For example in dichotomous choice, we can just set $\lambda=1$. If $\theta_i-\theta_j>0$, and $Y_{ij}^\alpha=-1$ so the residual $|\gamma_{ij}^\alpha|=|\theta_i-\theta_j-Y_{ij}^\alpha|>1$, then this comparison is easily to be picked out. On the other hand, if $Y_{ij}^\alpha=1$, $|\gamma_{ij}^\alpha|>1$ iff $\theta_i-\theta_j>2$, which is reasonable to be selected as an outlier.

\subsubsection{HLASSO Algorithm}
Based on these development, we have the the following Robust Ranking Algorithm \ref{alg:Outlier detection} called HLASSO here.
\begin{algorithm}
	\caption{Outlier Detection and Robust Ranking, denoted by HLASSO.}\label{alg:Outlier detection}
	{{\textbf{Initialization:} Compute the projection matrix $U_2$ via the SVD of $X$ or \eqref{eq:projection};}} \\ 
	\textbf{Solve the outlier detection LASSO \eqref{eq:HLASSO22}};\\
	\textbf{Tuning parameter}. Determine an optimal $\lambda^\ast$ by Huber's concomitant scale estimation (scaled-LASSO), or random projection based cross validation;\\
	\textbf{Rule out outlier effect and perform least squares in \eqref{eq:HLASSO21} to obtain the score estimation $\hat{\theta}$.}
\end{algorithm}

%

However, HLASSO suffers the following issues.
\begin{itemize}
	\item Bias: HLASSO gives a biased estimation \citep{FanLi01}, $\hat{\gamma}$ and $\hat{\theta}$.
	\item Scalability: \eqref{eq:HLASSO22} is prohibitive with large number of samples and ranking items.
\end{itemize}

In practice, one uses the subset least square estimate after dropping the outliers for a debiased estimator. In the following, we introduced a new algorithm based on Linearized Bregman Iterations which is both scalable and may remove bias in its limit.

\subsection{Linearized Bregman Iteration} \label{sec:LBI}
Here we introduce a new algorithm based on Linearized Bregman Iterations (LBI), which is a simple iteration scheme to minimize a loss function $L(\theta,\gamma)$ with sparsity control on $\gamma$:
\begin{subequations}
	\begin{align}
	\theta^{k+1} &= \theta^{k} - \triangle t \cdot \kappa \grad_\theta L(\theta^k,\gamma^k). \\
	z^{k+1} &= z^k - \triangle t \cdot \grad_\gamma L(\theta^k,\gamma^k).\\
	\gamma^{k+1}&=\kappa \Prox_{\|\cdot\|_1}(z).
	\end{align}
\end{subequations}
where the proximal map associated with a convex function $f:\R^n\to \R$ is given by $\Prox_{f}(z) :=\arg\min_x \frac{1}{2}\|x-z\|^2 + f(x)$. When $L(\theta,\gamma) = \|Y - X\theta -\gamma\|^2$, the algorithm is specified in Algorithm \ref{alg:LB1}.


\begin{algorithm}
	\caption{LBI in correspondence to (\ref{eq:HLasso})}\label{alg:LB1}
	\textbf{Initialization:} Given parameter $\kappa$ and $\triangle t$, define $h=\kappa \triangle t$, $k=0, z^0=0, \gamma^0=0$ and $\theta^0 = (X^TX)^{-1}XY$.\\
	\textbf{Iteration:}
	\begin{subequations}
		\begin{align}
		\theta^{k+1} &= \theta^{k} + h\,X^T (Y-X \theta^{k}-\gamma^{k}). \label{alg1Step1}\\
		z^{k+1} &= z^k + (Y-X \theta^{k}-\gamma^{k})\triangle t. \label{alg1Step2}\\
		\gamma^{k+1}&=\kappa\,\mathrm{shrink}(z^k). \label{alg1Step3}
		\end{align}
	\end{subequations}
	\textbf{Stopping:} exit when stopping rules are met. 
\end{algorithm}

%

\begin{remark}
	The parameter $h=\kappa \cdot \triangle t$ is the step size of gradient decent of $\theta$. So $h$ should not be too large to make the algorithm converge. In fact, we require that:
	\[h\|XX^T + I\| \le h (\|XX^T\| + 1) < 2,\] as shown in Section \ref{sec:consistency}.
\end{remark}

\begin{remark}
	The algorithm \ref{alg:LB1} generates a solution path of $\{t^k, \theta^k, \gamma^k\}_{k=1,\dots}, t^k = k\triangle t$, starting from a zero $\gamma^0$ and evolving into dense $\gamma^k$.  As we shall see in the next section, $t^k$ plays a similar role as $1/\lambda$ in (\ref{eq:HLasso}) to control the sparsity of $\gamma$. Sparsity is large as $k=0$ from the empty set, and decreases with possible oscillations as iteration goes. Early stopping regularization is thus required to avoid overfitting. Similar ways of parameter tuning as Huber-LASSO can be applied here to find an early stopping rule $\tau=k^\ast \triangle t$. 
\end{remark}

Moreover, one can change the update of $\theta^k$ from gradient decent in \eqref{alg1Step1} to exact solution: \[\theta^{k} = (X^TX)^{\dagger}X(Y-\gamma^{k}),\] which gives Algorithm \ref{alg:LB2}.

\begin{algorithm}
	\caption{LBI in correspondence to (\ref{eq:HLASSO22})}\label{alg:LB2}
	\textbf{Initialization:} Given parameter $\kappa,\triangle t$, define $k=0, z^0=0, \gamma^0=0$.\\
	\textbf{Iteration:}
	\begin{subequations}
		\begin{align}
		z^{k+1} &= z^k + (I-X(X^TX)^{\dagger}X^T)(Y-\gamma^{k})\triangle t. \label{alg2Step1}\\
		\gamma^{k+1}&=\kappa \,\mathrm{shrink}(z^k),\label{alg2Step2}
		\end{align}
	\end{subequations}
	\textbf{Stopping:} exit when stopping rules are met. 
\end{algorithm}

Note that $I-X(X^TX)^{\dagger}X^T$ is the projection matrix $U_2U_2^T$ which involves sparse matrix $X$ and is thus more efficient than finding $U_2$ in large scale problems. 

The algorithm \ref{alg:LB1} is also really easy to compute parallel, since it only needs matrix-vector multiplication.  Algorithm \ref{alg:LB-parallel} is the synchronized parallel version of Algorithm \ref{alg:LB1}.

\begin{algorithm}
	\caption{Sync-LBI of Algorithm~\ref{alg:LB1}}\label{alg:LB-parallel}
	\textbf{Initialization:} Given parameter $\kappa$, $\triangle t$ and thread number $P$, define $h=\kappa \triangle t$, $k=0, z^0=0, \gamma^0=0$ and $\theta^0 = 0$.\\
	\textbf{Split data and variables:} $\{1,\dots,m\} = \bigcup_{i=1}^P I_i$,~\,$\{1,\dots,n\} = \bigcup_{i=1}^P J_i$.\\
	\textbf{Iteration:} For each thread $i$
	\begin{subequations}
		\begin{align}
		res_{I_i} &= Y_{I_i}-X_{I_i} \theta^{k}-\gamma_{I_i}^{k}. \\
		z_{I_i}^{k+1} &= z_{I_i}^k + res_{I_i}\triangle t.\\
		\gamma_{I_i}^{k+1}&=\kappa\,\mathrm{shrink}(z_{I_i}^k).\\
		update^i &= X_{I_i}^Tres_{I_i}.
		\end{align}
	\end{subequations}
	Synchronize.
	\begin{equation}
	\theta_{J_i}^{k+1} = \theta_{J_i}^{k} + h\,\sum_{k=1}^P update^k_{J_i}.
	\end{equation}
	Synchronize.\\
	\textbf{Stopping:} exit when stopping rules are met. 
\end{algorithm}

\begin{figure}
	\begin{center}
		\subfigure[LASSO]{
			\includegraphics[width=0.48 \linewidth]{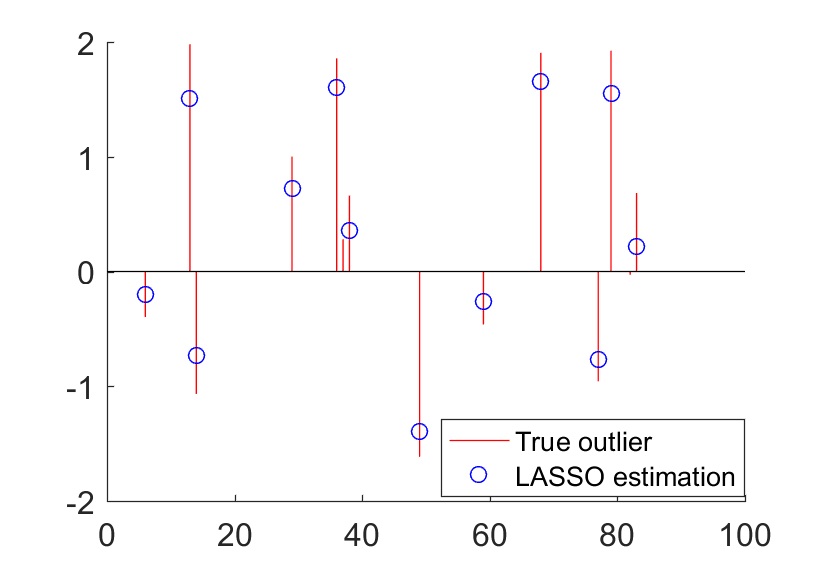}}
		\subfigure[LBI]{
			\includegraphics[width=0.44 \linewidth]{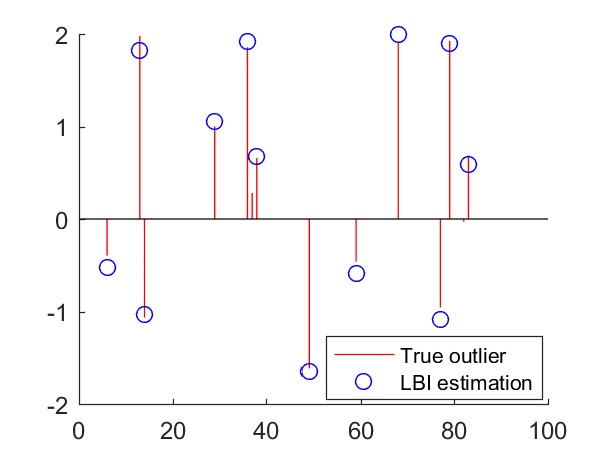}}
		\caption{Comparison of bias for LASSO and LBI with $\kappa=100$.}\label{fig:bias}
	\end{center}
\end{figure}

Despite its simplicity, LBI may reduce bias in its regularization path. Fig.\ref{fig:bias} shows a simple example to illustrate such a property of LBI. Here $n=10, m=100, \theta^\ast \sim U(0,1), \kappa=100, \epsilon\sim N(0,\sigma^2),\sigma = 0.1$. One can see the LBI estimator shows much smaller bias compared with that of LASSO. The error of LBI is mainly caused by Gaussian noise which also exists in the Oracle estimator. Actually it can be shown that the limit of LBI may find the Oracle estimator.

To see this, define $z^k = p^k + \frac{1}{\kappa}\gamma^k$ by Moreau Decomposition. When parameters $\kappa \to \infty$ and $\triangle t \to 0$, both Algorithm \ref{alg:LB1} and \ref{alg:LB2} converge to dynamics \eqref{eq:iss}, known as Inverse Scale Space (see \citep{osher2014} and references therein) describing the limit dynamics evolving from the null model to full:
\begin{subequations} \label{eq:iss}
	\begin{align}
	0 &= \nabla_{\theta} \|Y-X \theta-\gamma\|^2, \\
	\dot p  &= \nabla_{\gamma} \|Y-X \theta-\gamma\|^2,  \\
	p &\in \partial\|\gamma\|_1. \end{align}.
\end{subequations}

When sign-consistency is reached at certain $\tau$, i.e., $p(\tau)\in \partial \|\gamma^*\|_1$, then $\dot p_S(\tau)=0$ ($S=\supp(\gamma^*)$) and we have
\begin{eqnarray*}
	&&\left[
	\begin{array}{cc}
		X^T X & X_S^T \\
		X_S & I,
	\end{array}
	\right]
	\left[
	\begin{array}{c}
		\hat{\theta}(\tau) \\
		\hat{\gamma}_S(\tau)
	\end{array}
	\right]\\
	&=&
	\left[
	\begin{array}{cc}
		X^T X & X_S^T \\
		X_S & I
	\end{array}
	\right]
	\left[
	\begin{array}{c}
		\theta^\ast \\
		\gamma^\ast_S + \varepsilon_S
	\end{array}
	\right]
	+
	\left[
	\begin{array}{c}
		X_{S^c}^T \varepsilon_{S^c} \\
		0
	\end{array}
	\right].
\end{eqnarray*}
It implies that $E \hat{\gamma}_{S}(\tau) = \gamma_{S}^\ast$, i.e. the optimal estimator is unbiased.

\section{Identifiability of Outliers} \label{sec:consistency}
In this part, we show a statistical theory for outlier identifiability by HLASSO and LBI, respectively. As we shall see, the regularization path of LBI shares almost the same good properties as that of LASSO, i.e. outlier identifiability and $l_2$-error bounds of minimax optimal rates. They both consistently identify locations of outliers under nearly the same conditions. Such conditions, roughly speaking, require that the matrix $U_2$ satisfies an incoherence (or irrepresentable) condition that the locations of outliers are as independent as possible and the sparse outliers have large enough magnitudes. We shall see that large Erd\"{o}s-R\`{e}nyi random graphs meet the incoherence condition so that the random sampling in crowdsourcing may help outlier detection. Moreover, as the limit dynamics of LBI \eqref{eq:iss} is unbiased, LBI may estimate outlier magnitudes with less bias than HLASSO. Another distinction lies in that LBI exploits early stopping regularization while HLASSO chooses a regularization parameter $\lambda$.

Let $S =\rm supp (\ga^\ast)$ where $\supp(x) = \{i: x_i \neq 0\}$, $\tilde{y}=U_2^T y, \Psi = U_2^T$. $\Psi^T \Psi:l^2(E) \to l^2(E)$  is thus an orthogonal projection of $x\in l^2(E)$ to the kernel $\rm ker (X^T)$. Recall that the number of alternatives $n=|V|$ and the number of paired comparisons $m=|E|$. Denote the column vectors of $\Psi$ with index in $S$ ($S^c$) by $\Psi_S$ ($\Psi_{S^c}$) and the number of rows in $\Psi$ by $l$ ($l=m-n+1$). Let
\[ \mu_\Psi:= \max_{j\in S^c} \|\Psi_j\|_2^2,  ~\ga_{min} =\min_{i\in S} |\ga^\ast_i|,~s=|S|.\]

Now we have
\[ \tilde{y} = \Psi \ga^\ast + \tilde{\varepsilon} = \Psi_S \ga^\ast_S +  \Psi \varepsilon,\]
where $\varepsilon$ is sub-Gaussian with variance proxy $\sigma^2$.

\begin{asum}
	The following assumptions are required for both HLASSO and LBI.
	\begin{enumerate}
		\item[{C1}] Restricted Eigenvalue.
		\[ \Lambda_{\min} \left( \Psi_S^T \Psi_S \right) = C_{\min} > 0. \]
		\item[{C2}] Irrepresentability. For some constant $\eta\in (0,1]$,
		\[ \| \Psi^T_{S^c} \Psi_{S} (\Psi^T_{S} \Psi_{S})^{-1}  \|_\infty \leq 1 -\eta. \]
	\end{enumerate}
\end{asum}
Intuitively speaking these conditions suggests that the projection measurement of outliers in $\Psi$ should be as independent (orthogonal) as possible. In particular, {C1} suggests that $\Psi$'s columns associated with $S$ are highly independent; {C2} further implies that $\Psi_S$ contains distinct information from  $\Psi_{S^c}$ such that one can hardly represent the columns in $\Psi_{S^c}$ as elements in the column space of  $\Psi_S$.

\begin{figure}
	\begin{center}
		\subfigure [n=20]{
			\includegraphics[width=0.28 \linewidth]{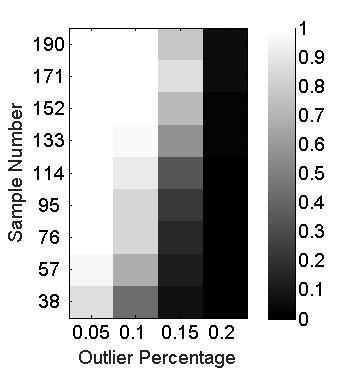}
		}
		\subfigure [n=60]{
			\includegraphics[width=0.28 \linewidth]{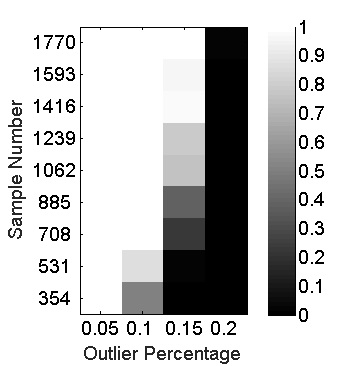}
		}
		\subfigure [n=100]{
			\includegraphics[width=0.28 \linewidth]{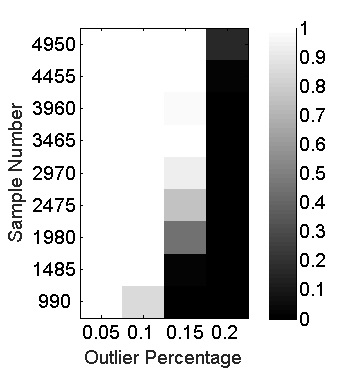} }
		\caption{The probability of C2 holding for uniform sampling. Every probability is estimated by 100 runs. White means 100\% and Black means 0\%.}\label{fig:C2}
	\end{center}
\end{figure}

The Irrepresentable condition C2, originally figured out by \citep{Tropp04,ZhaYu06} among others, is easy to satisfy for Erd\"{o}s-R\'{e}nyi random graphs. Fig.\ref{fig:C2} is a simple simulation to illustrate the probability that C2 holds under the uniform sampling generating Erd\"{o}s-R\'{e}nyi random graphs. One sees that when outlier percentage is not too large and sample number is sufficient, C2 holds with high probability. 

Asymptotically, a large Erd\"{o}s-R\'{e}nyi random graph needs $m\gg O(n\log n )$ to be connected, and $\Psi$ is the first $m-n+1$ rows of an orthogonal matrix. So $\Psi$ tends to be an orthogonal matrix as $n\to \infty$, because $n/m \ll n/(n\log n)\to 0$ where the correlation between rows of $\Psi$ diminishes. A sufficient condition by \citep{Donhuo01} (or see \citep{osher2014}) makes it easy to verify.

\begin{prop}
	Suppose (C0: Mutual Incoherence)
	$$\rho = \max_{i,j}X_i(X^TX)^{\dagger}X_j^T < \frac{1}{2s},$$
	then C1 and C2 hold with
	$$C_{min} = 1-s\rho,~\eta = \frac{1-2s\rho}{1-s\rho}.$$
\end{prop}
Note that $\Psi_T\Psi = I - X(X^TX)^{\dagger}X^T$, so C0 actually means $\Psi_T\Psi$ is close to Identity matrix.
\begin{remark}
	Let $L = E(X_iX_i^T)$, which is only determined by the sampling method. Moreover, as $m$ goes to infinity, $X^TX/m\rightarrow L$. Then $\rho \le 4\max_{i,j}(X^TX)^{\dagger}\approx \frac{4}{m}\max_{i,j}L^{\dagger}$. For uniform sampling, $L = \frac{2}{n(n-1)}(nI - 11^T)$, thus $L^{\dagger} = \frac{n-1}{2}(I - 11^T/n)$, $\rho \le \frac{2n}{m}$. So as long as $\frac{2n}{m} < \frac{1}{2s}$, C1 and C2 hold with high probability for uniform sampling. 
\end{remark}

Equipped with C1 and C2, we then provide theoretical analysis for HLASSO and LBI from the following three aspects:
\begin{enumerate}
	\item[(A1)] \textit{No-false-positive}. We will show that, with  high probability,  $supp(\hat{\gamma}) \subseteq supp(\gamma^*)$, i.e.,  all the detected outliers are true outliers under the assumptions (zero false-positive-rate).
	\item[(A2)] \textit{Sign Consistency}. With an extra assumption that outliers are strong enough, we can actually show that $supp(\hat{\gamma}) = supp(\gamma^*)$ holds with high probability, which is an even stronger result than (A1) implying the detected outliers  are consistent with the true outliers.
	\item[(A3)] \textit{$\ell_2$ bound}. It could also be proved that the  $\ell_2$ distance between $\hat{\gamma}$ and $\gamma^*$ is bounded above by the same rates $O(\sqrt{s \log m})$ that is statistically optimal. In this way  $\hat{\gamma}$ could also capture the magnitude of its ground-truth $\gamma^*$.
\end{enumerate}

The following theorem extends \citep{Wainwright09} to the specific setting of HLASSO in this paper.

\begin{theorem}[Outlier Identifiability of HLASSO]\label{thm:LASSO}
	Let
	\[ \overline{\lambda} = \frac{2\sigma\sqrt{\mu_\Psi}}{\eta} \sqrt{\log m}.\]
	Then with probability greater than $1-2/m$,
	the unique solution path of (\ref{eq:HLASSO22}) $\hat{\ga}_\lambda$ satisfies:
	\begin{itemize}
		\item \textbf{(No-false-positive)} If C1 and C2 hold, for all $\lambda \ge \overline{\lambda}$, there are no false positive outliers, i.e., $\supp(\hat{\gamma}_\lambda)\subseteq S$.
		\item \textbf{(Sign-consistency)} If C1 and C2 hold, and for $c_0 =  \frac{\eta }{\sqrt{C_{\min} \mu_\Psi} } + \|(\Psi^T_{S} \Psi_{S})^{-1}  \|_\infty$ assume
		\begin{equation}\label{eq:C3}
		\mbox{(C3: Large Outlier) } \ga_{min} > c_0\overline{\la}=O(\sqrt{\log m}),
		\end{equation}
		then for $\lambda\in [ \overline{\lambda} , \ga_{min}/c_0)$, there will be no false positive and negative outliers, i.e., $\supp(\hat{\gamma}_\lambda) = S(\ga^\star)$ ($\sign(\hat{\ga}_\la) = \sign(\ga^\ast)$.
		\item \textbf{($\ell_{2}$-bound)} If C1 and C2 hold, then for all $\lambda \ge \overline{\lambda}$,
		\[\| \hat{\ga}_{\la} - \ga^\star_S \|_2 \leq c_0\sqrt{s}\lambda .\]
	\end{itemize}
\end{theorem}

\begin{remark}
	We have the following remarks for Theorem \ref{thm:LASSO}:
	\begin{itemize}
		\item Condition C1 is necessary for the uniqueness of a sparse outlier $\ga^\ast$ whose support is contained in $S$. C2 and C3 are sufficient and necessary to ensure no-false-positive outliers and no-false-negative outliers, respectively, as discussed in \citep{Wainwright09}.
		\item  $\| \hat{\ga}_{\overline{\la}} - \ga^\ast_S \|_2 \leq  O(\sqrt{s \log m})$, which is the minimax-optimal rate up to a logarithmic factor. Such a rate can be achieved under weaker assumptions without ensuring the model selection consistency \citep{bickel2009}.
	\end{itemize}

\end{remark}

The next theorem extends \citep{osher2014} to our specific setting, showing that under nearly the same conditions, LBI can achieve similar results.

\begin{theorem}[Outlier Identifiability of LBI]\label{thm:LB}
	Let $h=\kappa \triangle t$, $B = \gamma_{max} + 2 \sigma \sqrt{\frac{\log m}{C_{min}}}  + \frac{\|\Psi\gamma\|_2 + 2s\sqrt{\log l}}{\sqrt{C_{min}l}}$, and
	\[\overline{\tau} = \frac{(1 - B/\kappa\eta)\eta} {2 \sigma\sqrt{\mu_\Psi}} \sqrt{\frac{1}{\log m}} = (1 - \frac{B}{\kappa\eta})\frac{1}{\overline{\lambda}}.  \]
	Assume $\kappa$ is big enough to satisfy $B\le\kappa\eta$, and step size $\triangle t$ is small s.t. $h\|\Psi_S \Psi_S^T\|<2$. Then with probability at least $1-\frac{2}{m}-\frac{2}{l}$, Algorithm \ref{alg:LB2} has paths satisfying:
	\begin{itemize}
		\item \textbf{(No-false-positive)} If C1 and C2 hold, for all $k$ such that $t^k\leq \overline{\tau}$, the path has no-false-positive, i.e., $\supp(\gamma^k)\subseteq S$;
		\item \textbf{(Sign-consistency)} If C1 and C2 hold, and moreover, if the smallest magnitude $\gamma_{min}$ is strong enough and $\kappa$ is big enough to ensure
		\[\gamma_{\min} \geq \frac{4 \sigma}{C_{min}^{1/2}}\sqrt{\log m} ,\]
		\begin{equation}\label{C3}
		\overline{\tau} \geq \frac{8+4\log{s}}{\tilde{C}_{min}\gamma_{min}}+\frac{1}{\kappa \tilde{C}_{min}}\log(\frac{3\|\gamma\|_{2}}{\gamma_{\min}}) + 3\triangle t,
		\end{equation}
		where $\tilde{C}_{min} = C_{min} (1- h\|\Psi_S \Psi_S^T\|/2)$, then for $k^*=\lfloor\bar{\tau}/(\triangle t)\rfloor$, $\sign(\gamma^{k^*})=\sign(\gamma)$.
		\item \textbf{($\ell_{2}$-bound)}. If C1 and C2 hold, for some constant $\kappa$ and $C$ large enough such that
		\begin{align*}
		\bar{\tau}\ge &\frac{1}{2\kappa \tilde{C}_{min}}(1+\log{\frac{\|\gamma\|^{2}_{2}+4\sigma^2s\log{m}/C_{min}}{C^{2}s\log{m}}}) \ldots \\
		&\ldots +\frac{4}{C\tilde{C}_{min}}\sqrt{\frac{1}{\log m}} + 2\triangle t,
		\end{align*}
		there is a $k^*$, $t^{k^*}\leq \bar{\tau}$, such that
		$$\|\gamma^{k^*}-\gamma\|_{2} \leq (C+\frac{2\sigma}{C_{min}^{1/2}}) \sqrt{s\log m}.$$
	\end{itemize}
\end{theorem}

\begin{remark} We have two final remarks on the theoretical analysis:
	\begin{itemize}
		\item The theorems show that $t^k$ in Algorithm \ref{alg:LB2} plays the same role of $1/\lambda$ in (\ref{eq:HLASSO22}),
		\[ \overline{\tau}\overline{\lambda} = (1 - B/\kappa\eta) \rightarrow 1,~\mbox{as}~\kappa\rightarrow\infty. \]
		Therefore, early stopping in LBI plays a role of regularization as tuning $\lambda$ in LASSO.
		\item The condition \eqref{C3} requires $\ga_{min} > O(\sqrt{\log m} \log s)$, which is slightly stronger than the condition C3 in \eqref{eq:C3} up to a $\log s$ factor.
	\end{itemize}
\end{remark}

\section{Experiments}\label{sec:experiments}

In this section, 7 datasets are exhibited with both
simulated and real-world data to illustrate the validity
of the analysis above and applications of the methodology
proposed. The first two examples are with
simulated data while the latter 5  exploit real-world
data in VQA, IQA, human age predication, face verification and fine-grained attribute retrieval for shoes.

\subsection{Simulated Study} \label{sec:simulatedata}
\subsubsection{Experiment I: Synthetic pairwise data}

\begin{figure}
	\begin{center}
		\subfigure[LASSO]{
			\includegraphics[width=0.48\columnwidth]{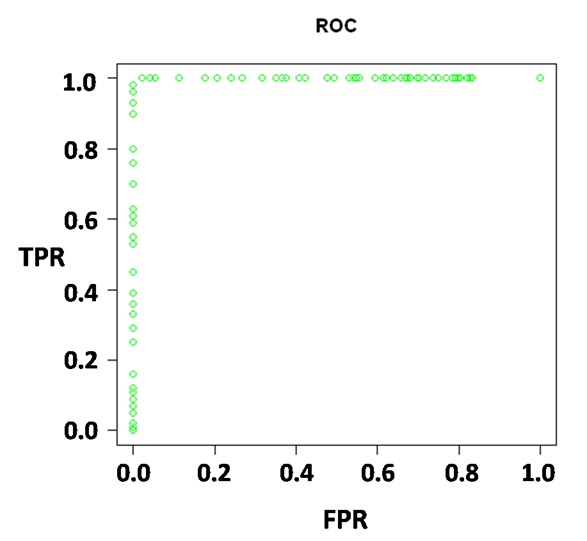}\label{ROC}}
		\subfigure[LBI]{
			\includegraphics[width=0.48\columnwidth]{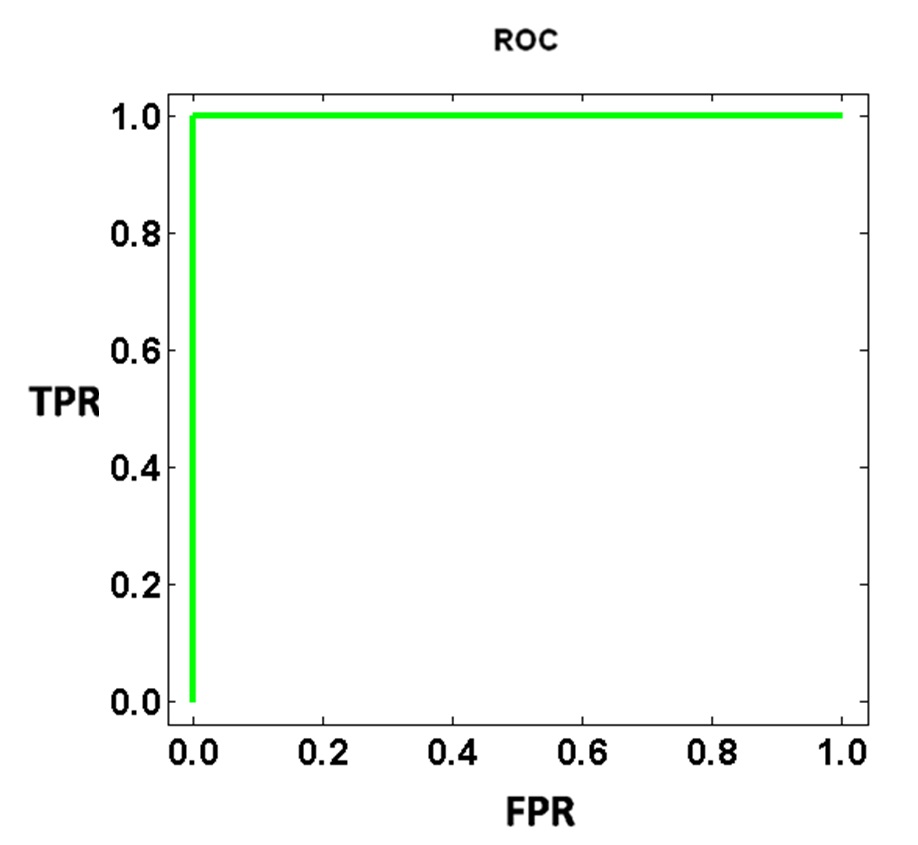}\label{ROC2}}
		\caption{ROC curve of ($2000,5\%$) of simulated data.}
	\end{center}
\end{figure}

\begin{figure}
	\begin{center}
		\includegraphics[width=0.95\linewidth]{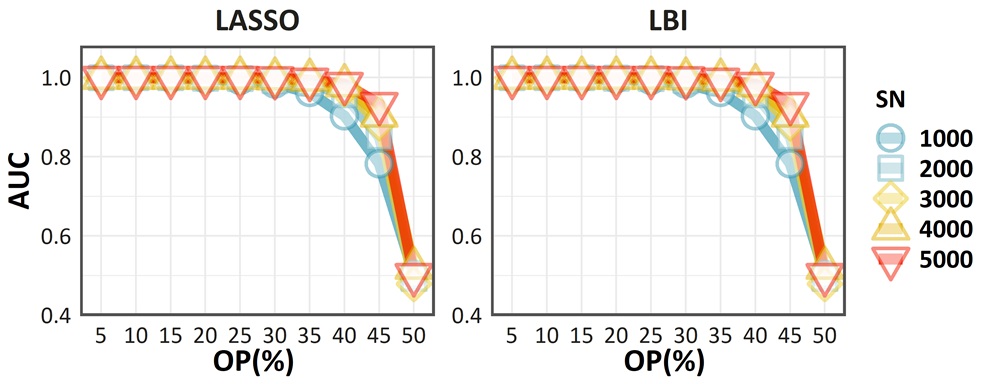}
		\caption{AUC performance curves with various Sample Number (SN) and Outlier Percentage (OP) on the simulated data. The points in the figure are obtained from the mean value over 20 repetitions. } \label{fig:auc_sim}
	\end{center}
\end{figure}

In this experiment, we first create a random total order on $n$ candidates $V$ as the
ground-truth and add paired comparison edges $(i,j)\in E$ to graph $G=(V,E)$ randomly, with the
preference direction following the ground-truth order. To create sparse outliers,
a random subset of $E$ is reversed in preference direction. In this way,
we simulate a paired comparison graph, possibly incomplete and imbalanced, with outliers.

Here we choose $n=|V|=16$, which is consistent with the first two real-world datasets. For convenience, denote the total number of paired comparisons by SN (Sample Number), the number of outliers by ON (Outlier Number), and the outlier percentage by OP = ON/SN. The following will show the proposed LBI could exhibit comparable performance with LASSO for outlier detection.

\begin{figure}
	\begin{center}
		\subfigure[LASSO]{
			\includegraphics[width=0.48 \linewidth]{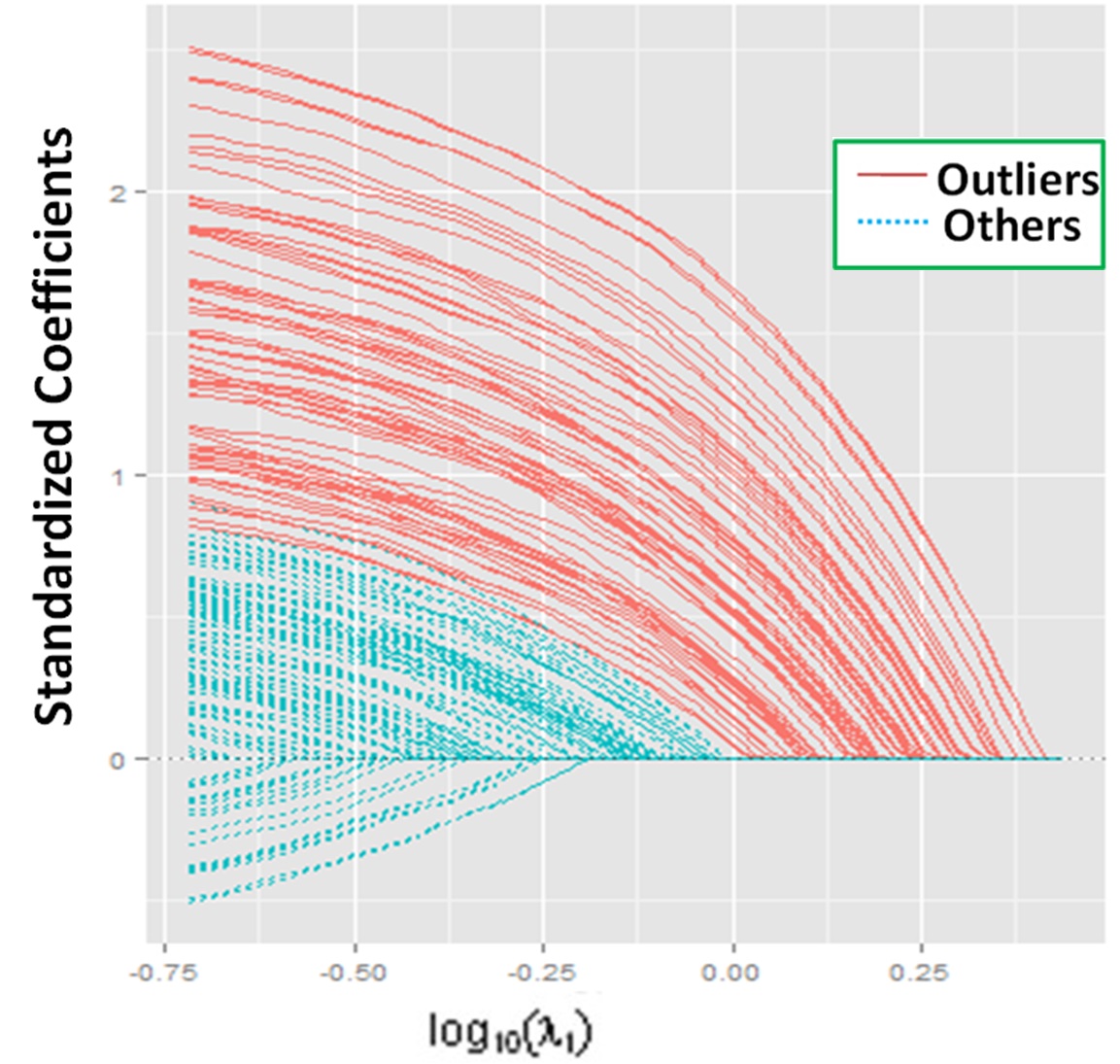}\label{pathlasso}}
		\subfigure[LBI]{
			\includegraphics[width=0.48 \linewidth]{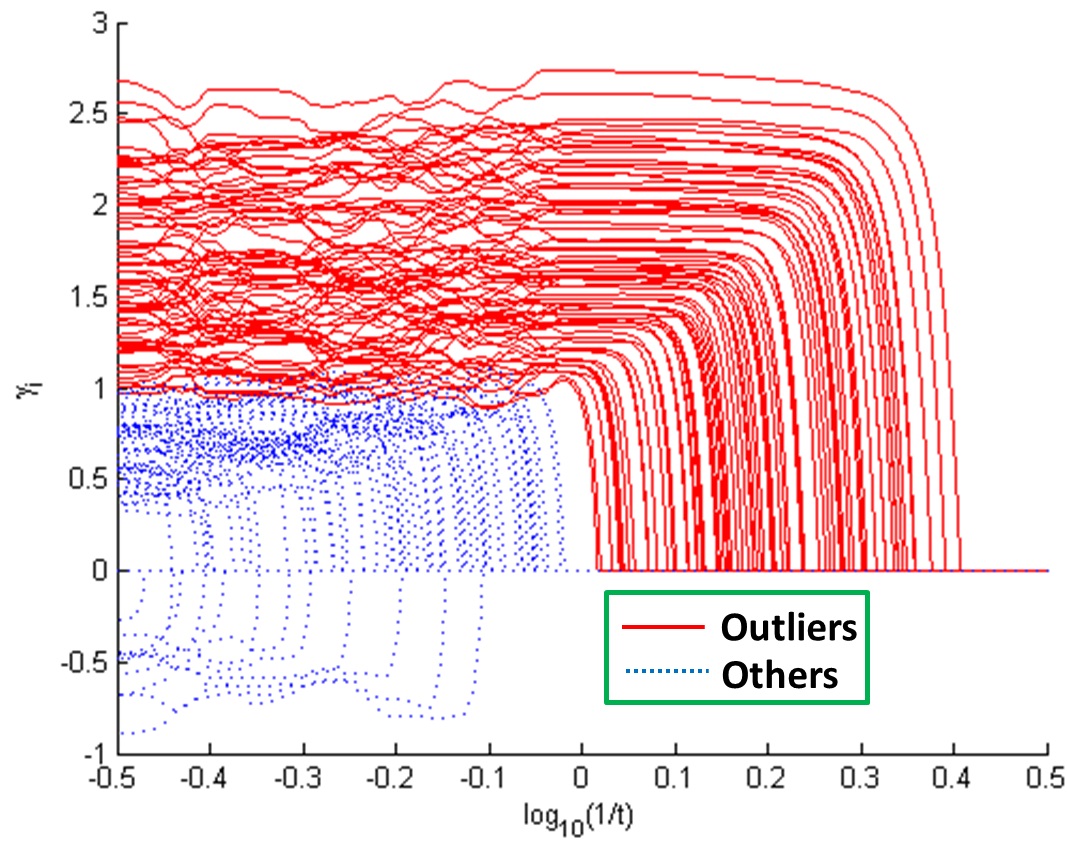}\label{pathlbi}}
		\caption{The regularization paths of ($2000,5\%$) of LASSO vs. LBI. The true outliers (plotted in red color) mostly lie outside the majority of paths.}
	\end{center}
\end{figure}

First, for each pair of (SN,OP), we compute the regularization path $\hat{\gamma}^\lambda$ of LASSO by varying regularization parameter $\lambda$ from $\infty$  to
$0$, which is solved by R-package {\texttt{quadrupen}} \citep{quadrupen}. The order in which $\hat{\gamma}_{ij}^{\lambda}$ becomes nonzero gives a ranking of the edges according to
their tendency to be outliers. Since we have the ground-truth outliers, the ROC curve can be plotted by thresholding the regularization
parameter $\lambda$ at different levels which creates different true positive rates (TPR) and false positive
rates (FPR). For example, when SN = 2000 and OP = 5\%, the ROC curve can
be seen in Fig.\ref{ROC}. With different choices of SN and OP, Area Under the Curve (AUC) are computed with the average result over 20 runs and shown in Fig.\ref{fig:auc_sim} to measure the
performance of LASSO in outlier detection. Moreover, comparable results returned by LBI can be found in Fig.\ref{ROC2} and Fig.\ref{fig:auc_sim}. It can be seen that when samples are large and outliers are sparse, the AUC of both of these two methods
are close to 1. This implies that both LASSO and LBI can provide an accurate estimation of
outliers (indicated by a small FPR with large TPR). Fig.\ref{pathlasso}, \ref{pathlbi} illustrate the regularization path examples of LASSO and LBI.

We note that when OP = 50\%, i.e., half of the edges are reverted by outliers, Fig.\ref{fig:auc_sim} shows a rapid decrease of AUC to about 0.5, which is the performance of random guess. This is as expected, since when more than half of the edges are perturbed, it is impossible to distinguish the signal from noise by any method. A phase transition can be observed in figure, that AUC rapidly increases to 1 as long as OP drops below 50\% and SN increases.

The simulated example mentioned above tells us that LBI could exhibit similar performance with LASSO in most cases when sample numbers are not large. But when the sample number grows to be large, LASSO paths will be too expensive to compute while LBI still scales up. The following experiment provides such an example.

\subsubsection{Experiment II: Image Reconstruction}
In some scenarios like image reconstruction, there is a large number of samples (paired comparisons between pixels in images) and every sample contributes to an outlier variable, which makes LASSO difficult to detect outliers effectively. However LBI could provide us a simple yet scalable outlier detection algorithm for large-scale data analysis.

\begin{figure}[tp]
	\begin{center}
		\includegraphics[width=0.95\linewidth]{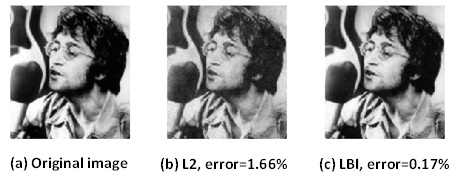}
		\caption{Results of L2/LBI for image reconstruction with mean squared error.} \label{image_recons}
	\end{center}
\end{figure}

Here we use a simulation example of image reconstruction from \citep{Yu12}. First we use an image as the ground-truth, with an intensity range of $[0,1]$ over $181\times 162$ pixels. Local comparisons are obtained as intensity differences between pixels within a $5\times 5$ neighborhood, added with Gaussian noise of $\sigma=0.05$. Furthermore 10\% of these measurements are added with random outliers of $\pm 0.5$. So there are $n=29,322$ nodes and $m=346,737$ pairwise comparisons. In this example, none of the LASSO packages in this paper can deal with this example; in contrast, LBI takes the advantage of sparsity of $X$ and works well. Fig.\ref{image_recons} shows the results marked
by their mean squared errors with respect to the original image where LBI exhibits a significantly smaller error than the least squares (L2).

\begin{figure}[tp]
	\begin{center}
		\subfigure[n=1000]{
			\includegraphics[width=0.48 \linewidth]{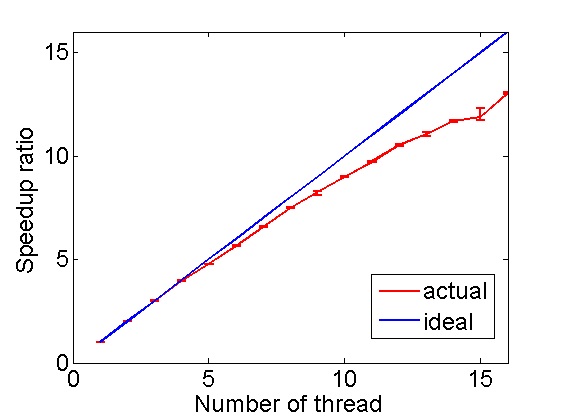}\label{1000lbi}}
		\subfigure[n=2000]{
			\includegraphics[width=0.48 \linewidth]{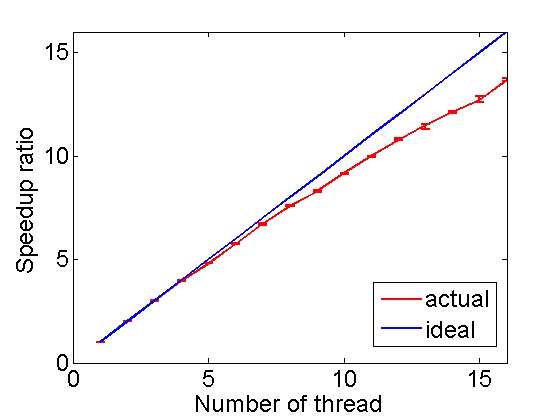}\label{2000lbi}}
		\caption{The speedup of parallel LBI.}
	\end{center}
\end{figure}

\subsubsection{Experiment IV: Speedup of parallel LBI}
In this experiment, we demonstrate the linear speedup of parallel LBI for large-scale dataset. Similar to the experiment setting in Experiment I, we choose $n=|V|=1000$, $SN=10*n^2$ and $OP=20\%$, in this experiment. In evaluating a parallel system, one typical performance measure
is speedup, which is
defined as the ratio of the
elapsed time when executing a program on a single thread
(the single thread execution time) to the execution time
when $N$ threads are available. Let $T(N)$ be the time required to complete the task on $N$ threads. The speedup $S(N)$ is the ratio:
\begin{equation}
S(N)=\frac{T(1)}{T(N)}
\end{equation}

In our setting, $N=1, 2, 3,...16$. Fig.\ref{1000lbi} shows the error bar of speedup with confidence interval [0.25 0.75] for 20 times repeat of Parallel LBI in a 16-core server, with Intel(R) Xeon(R) E5-2670 2.60GHz CPU and 384GB of RAM. The server runs Linux 4.2.0 64bit. It is easy to find that the parallel LBI could speed up the running time almost in a linear manner. Moreover, similar phenomenon happens for $n=2000$ in Fig.\ref{2000lbi} which further demonstrates the effectiveness of our proposed parallel LBI.

\subsection{Real-world Datasets}

In this section, we will proceed to see the performance of our proposed algorithms on real-world datasets. The experiments are carried out on a wide range of datasets including three continuous ranking datasets for image/video quality assessment and age ranking, two bipartite ranking datasets for face verification and fine-grained retrieval.

During this subsection, we employ several competitors in the experimental analysis.

\begin{enumerate}
\item \textbf{Ordinary Rankers}. To show whether the outlier detection methods are effective in removing the redundant information from the dataset, we first adopt some popular ordinary ranking models without outlier detection mechanism as standard baselines.
\begin{itemize}
\item[$\star$] \textbf{The Least Squared Method (L2)}. This is the simplest method we could use in crowdsourced ranking. To implement this method, one only needs to solve the L2 minimization problem $\min_{s} \sum_{(i,j)} (y_{ij} - (s_i - s_j))^2$.
\item[$\star$] \textbf{eXtreme Gradient Boosting Linear model (XGB\_Linear)}. We also include the stronger baseline XGB  \cite{xgboost}, which is recently a famous choice for a lot of winners in the Kaggle competitions. Here we implement the Linear model version to gain the ranking score.
\item[$\star$] \textbf{Principle Component Guided Lasso (PC-\\LASSO)}. This is a recently proposed model \cite{PCLASSO}, which provides a strengthened sparsity-leveraging penalty with the principal component directions in the feature space.
\end{itemize}

\item \textbf{Robust Rankers} We also provide three robust competitors summarized as follows.
\begin{itemize}
\item[$\star$] \textbf{iHT}. This is a $\ell_0$-norm-based robust outlier detection method proposed in \cite{outxu}. Specifically, it solves the following minimization problem:
\begin{equation}
\begin{split}
\min_{{s}, {E}} & \frac{1}{2} ||{Y} - {X}{s} -{E}||_2^2 \\
& s.t.~~ ||{E}||_0 \le K,
\end{split}
\end{equation}
where the $\ell_0$-norm constrained sparse vector $E$ captures the outliers that are conflict with the global ranking.

\item[$\star$] \textbf{iLTS}. This algorithm  \cite{outxu} provides an alternative way to perform $\ell_0$-norm-based robust outlier detection.  Specifically, it solves the following minimization problem:
\begin{equation}
\begin{split}
\min_{{s}, {\Lambda}} & \frac{1}{2} ||\Lambda \odot ({Y} - {X}{s})||_2^2 \\
& s.t.~~ \Lambda \in \{0,1\}^N, ||\Lambda||_0 \ge N-K,
\end{split}
\end{equation}
where $\Lambda$ provides the indicator for the normal instances. The outliers are then regarded as the samples corresponding to the zero-entries of $\Lambda$.
\item[$\star$] \textbf{LASSO+L2}. It is well-known that the sparse penalties often lead to an unbiased estimator of the ranking score. To leverage an unbiased estimator, we adopt a two-stage way to learn a robust score. First, we employ the HLASSO model to identify the outliers and then remove the outliers that the model gives strong confidence. Then, we perform an unbiased L2 method again on the clean data to obtain the final scores.
\end{itemize}

\end{enumerate}

As there is no ground-truth for outliers in real-world data, one cannot exploit ROC and AUC as in simulated data to evaluate outlier detection performance here.
In this subsection, we inspect the top $p\%$ pairs returned by LASSO/LBI and compare them with the whole data to see
if they are reasonably good outliers. Besides, to see the effect of outliers on global ranking scores, we will further show the results from the 6 competitors introduced above together with HLASSO and LBI. Since LASSO+L2 is an unbiased algorithm, we will also see how others fit with this algorithm. Overall, numerical experimental results fit our theory nicely. Besides, we denote LBI algorithms by LBI ($\kappa,\triangle t$) for parameter choices.

\begin{table*}[h]
\caption{Comparison of different rankings on PC-VQA data. Eight algorithms are compared here with the integer representing the ranking position and the number in parenthesis representing the global ranking score returned by the corresponding algorithm. The first column provides the ID of the instances ordered by the score from the LASSO+L2 algorithm. The items in red are the instances that are inconsistent with the ranking results of LASSO+L2. }\label{tab:data5-rank}
\centering
    \begin{tabular}{lcccccccc}
     ID     & LASSO+L2 & L2    & XGB\_Linear & PCLASSO & iLTS  & iHT   & HLASSO & LBI \\
    \toprule
    1     & 1 ( 0.8688) & 1 ( 0.7930) & 1 ( 0.7914) & 1 ( 1.0692) & 1 ( 0.9121) & 1 ( 0.8746) & 1 ( 0.8103) & 1 ( 0.8648) \\
    9     & 2 ( 0.5996) & 2 ( 0.5313) & 2 ( 0.5302) & 2 ( 0.8910) & 2 ( 0.6467) & 2 ( 0.5996) & 2 ( 0.5478) & 2 ( 0.5987) \\
    10    & 3 ( 0.5253) & 3 ( 0.4805) & 3 ( 0.4795) & 3 ( 0.8240) & 3 ( 0.5501) & 3 ( 0.5254) & 3 ( 0.4892) & 3 ( 0.5243) \\
    13    & 4 ( 0.5100) & 4 ( 0.3906) & 4 ( 0.3899) & 4 ( 0.7931) & 4 ( 0.5201) & 4 ( 0.5100) & 4 ( 0.4155) & 4 ( 0.5059) \\
    7     & 5 ( 0.4570) & 5 ( 0.2852) & 5 ( 0.2846) & 5 ( 0.7258) & 5 ( 0.5132) & 5 ( 0.4510) & 5 ( 0.3104) & 5 ( 0.4266) \\
    8     & 6 ( 0.3156) & 6 ( 0.2383) & 6 ( 0.2378) & \textcolor{red}{8 ( 0.5646)} & 6 ( 0.3391) & 6 ( 0.3156) & 6 ( 0.2501) & 6 ( 0.3059) \\
    11    & 7 ( 0.2601) & 7 ( 0.2148) & 7 ( 0.2144) & \textcolor{red}{6 ( 0.6030)} & 7 ( 0.2823) & 7 ( 0.2601) & 7 ( 0.2234) & 7 ( 0.2550) \\
    14    & 8 ( 0.2125) & 8 ( 0.1641) & 8 ( 0.1637) & \textcolor{red}{7 ( 0.5847)} & 8 ( 0.1898) & 8 ( 0.2124) & 8 ( 0.1719) & 8 ( 0.2061) \\
    15    & 9 (-0.1749) & 9 (-0.1758) & 9 (-0.1754) & \textcolor{red}{10( 0.1920)} & 9 (-0.1808) & 9 (-0.1750) & 9 (-0.1785) & 9 (-0.1817) \\
    12    & 10(-0.2800) & \textcolor{red}{11(-0.2500)} & \textcolor{red}{11(-0.2495)} & \textcolor{red}{11( 0.1475)} & 10(-0.2921) & 10(-0.2800) & \textcolor{red}{11(-0.2562)} & 10(-0.2781) \\
    3     & 11(-0.3017) & \textcolor{red}{10(-0.2227)} & \textcolor{red}{10(-0.2222)} & \textcolor{red}{9 ( 0.3055)} & 11(-0.3553) & 11(-0.3049) & \textcolor{red}{10(-0.2361)} & 11(-0.2918) \\
    4     & 12(-0.3608) & 12(-0.2930) & 12(-0.2924) & \textcolor{red}{14( 0.0879)} & 12(-0.3567) & 12(-0.3667) & 12(-0.3015) & 12(-0.3498) \\
    16    & 13(-0.4812) & 13(-0.3633) & 13(-0.3626) & \textcolor{red}{12( 0.1405)} & 13(-0.4652) & 13(-0.4810) & 13(-0.3788) & 13(-0.4673) \\
    5     & 14(-0.5760) & 14(-0.4414) & 14(-0.4405) & \textcolor{red}{13( 0.0947)} & 14(-0.6209) & 14(-0.5760) & 14(-0.4651) & 14(-0.5703) \\
    6     & 15(-0.7412) & 15(-0.6289) & 15(-0.6277) & 15(-0.0846) & 15(-0.8180) & 15(-0.7412) & 15(-0.6570) & 15(-0.7398) \\
    2     & 16(-0.8332) & 16(-0.7227) & 16(-0.7212) & 16(-0.2195) & 16(-0.8645) & 16(-0.8239) & 16(-0.7455) & 16(-0.8086) \\
    \bottomrule
    \end{tabular}%
\end{table*}

\begin{figure}
	\setlength{\abovecaptionskip}{5pt}
	\begin{center}
		\includegraphics[width=\linewidth]{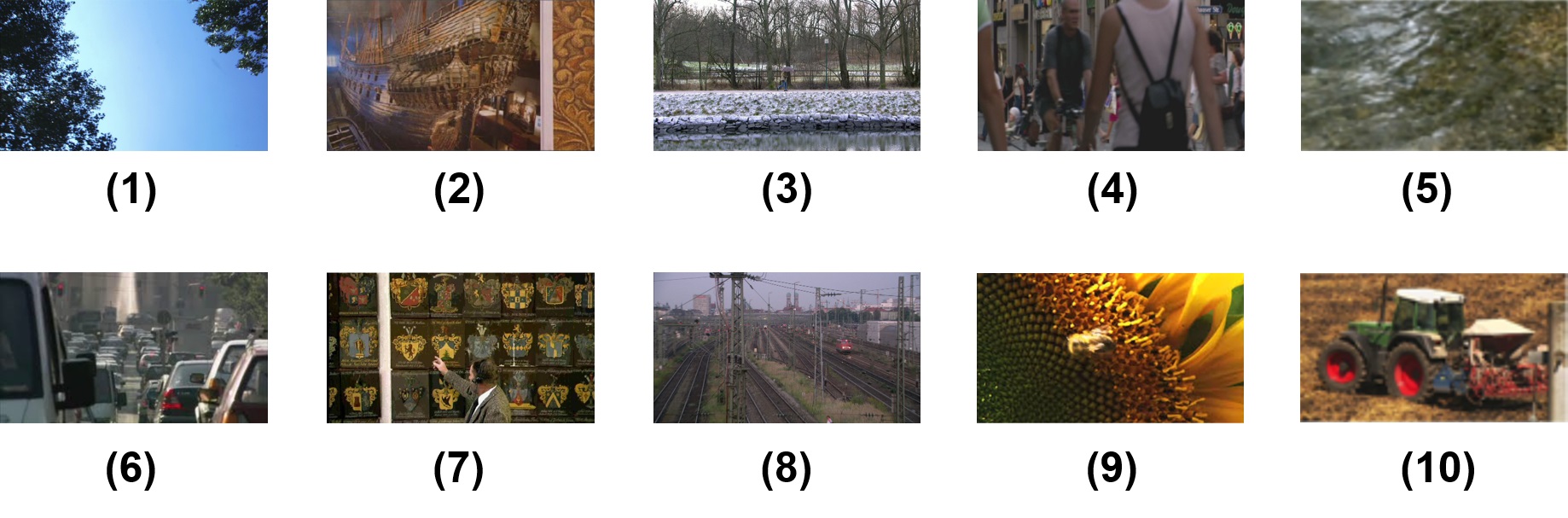}
		\caption{Reference videos in LIVE database.} \label{datavideos}
	\end{center}
\end{figure}

\begin{figure}
\begin{center}
	\includegraphics[width=\linewidth]{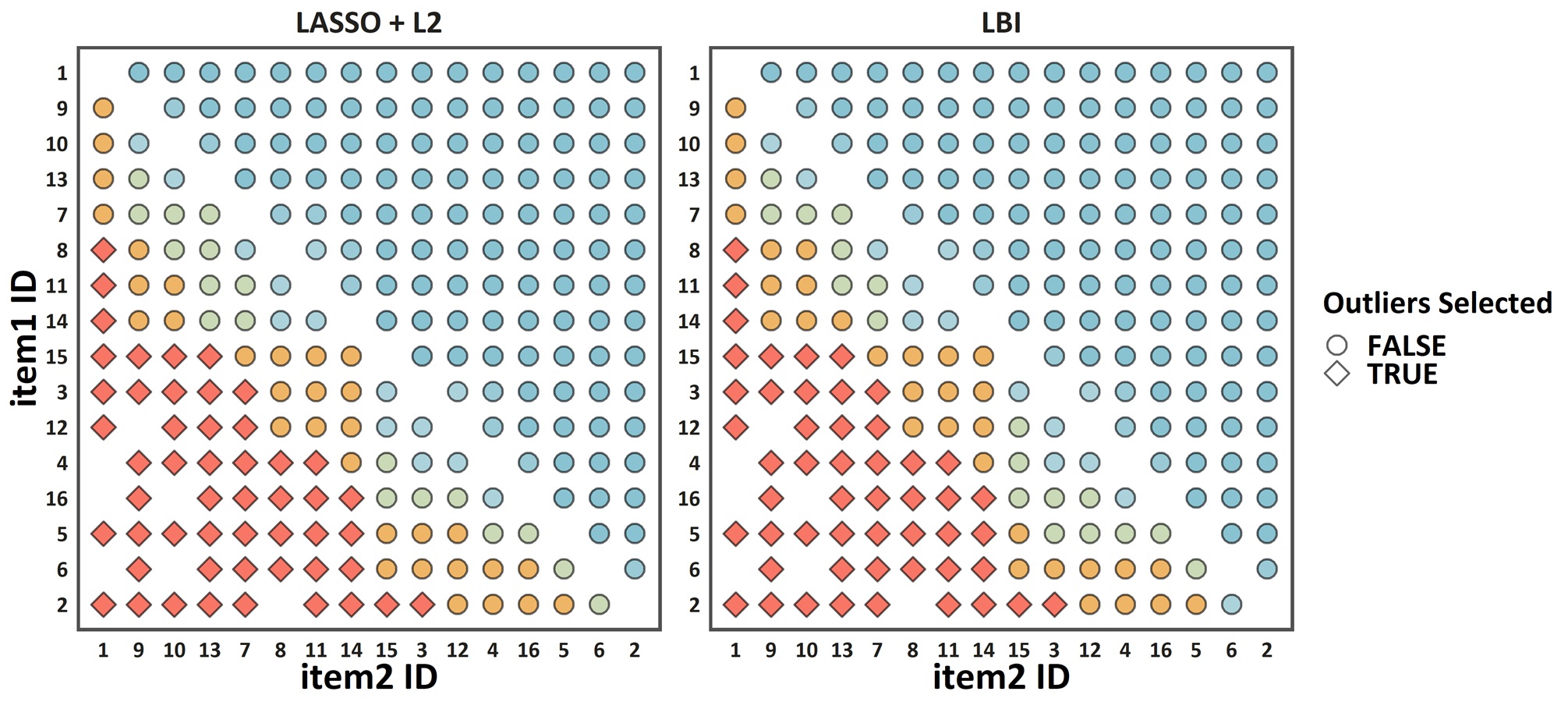}
	\caption{Visualization of the outliers in the PC-VQA dataset. The shape of points represents whether the given pair is selected as an outlier by the underlying algorithm.  Each point here represents a comparison pair (item1, item2), the $x$-axis gives the ID for item2 while the $y$-axis gives the ID with item1. The labels in the $x$- and $y$-axis are arranged according to the score order from the L2 method (ascending). The color of points represents the strength of outliers. We adopt a color gradient from red to blue to represent the top 5\%, 10\%, 15\%, 20\%, 25\% outliers and the other pairs.    } \label{fig:diffvqa}
\end{center}
\end{figure}

\begin{figure*}
	\begin{center}
		\includegraphics[width=0.75\linewidth]{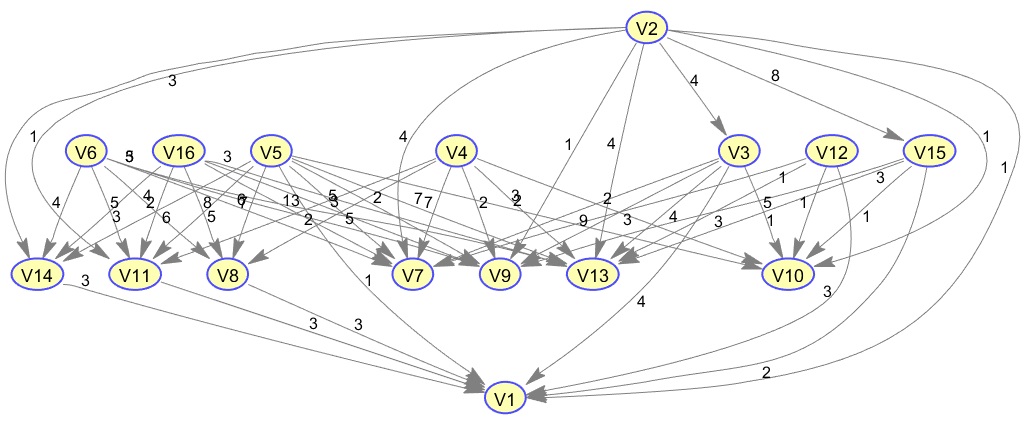}
		\caption{Top 5\% outliers  in PC-VQA dataset.} \label{data5}
	\end{center}
\end{figure*}

\begin{figure}
	\begin{center}
		\includegraphics[width=\linewidth]{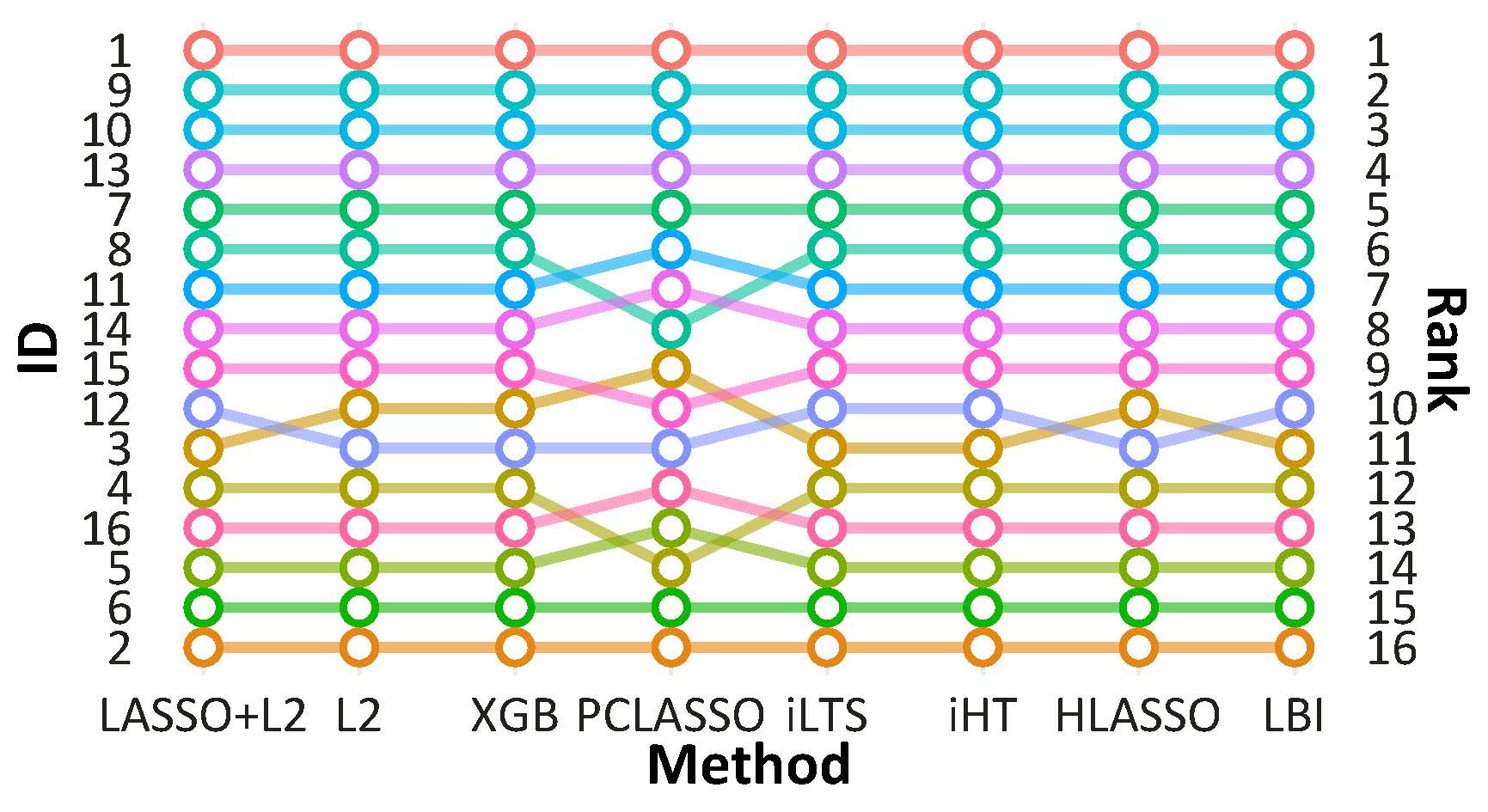}
		\caption{Visualization of the Rank on PC-VQA dataset with bump chart. Each line here tracks the rank of a given item across different algorithms. There is a change in rank whenever a line inclines to the x-axis. } \label{fig:vqa_rank}
	\end{center}
\end{figure}

\subsubsection{PC-VQA Dataset} \label{sec:VQA}
The first dataset, PC-VQA, collected by \citep{MM11}, contains $38,400$ paired comparisons of the LIVE dataset \citep{LIVE} (as shown in Fig.\ref{datavideos}) from 209 random observers. An attractive property of this dataset is that the paired comparison data is complete and balanced. As LIVE includes 10 different reference videos and 15 distorted versions of each reference,
for a total of 160 videos, the complete comparisons of this video database
require  $10 \times { 16 \choose 2 }  =1200 $ comparisons. Therefore, $38,400$ comparisons correspond to 32 complete rounds.

	\begin{figure}
	\begin{center}
		\includegraphics[width=\linewidth]{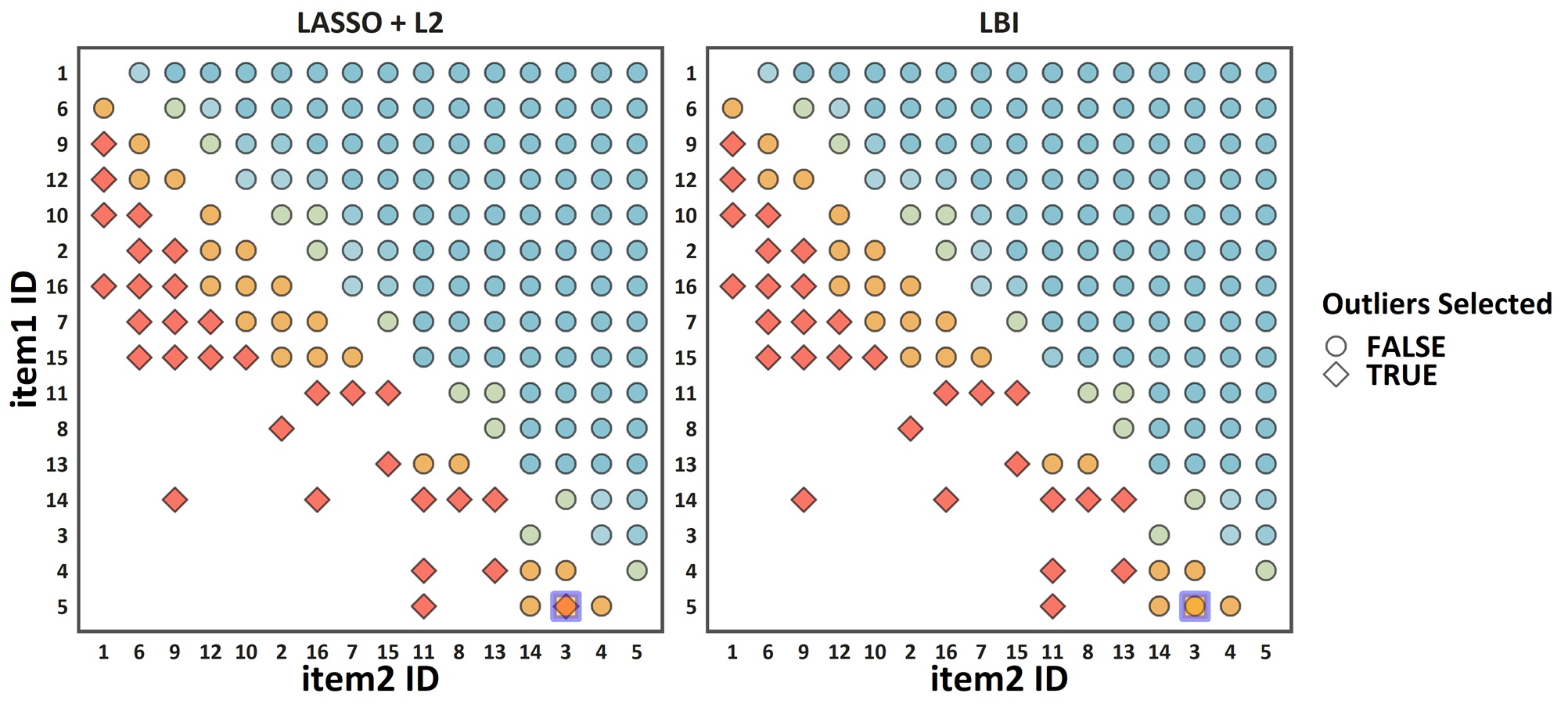}
		\caption{Visualization of the outliers in the IQA dataset. The shape of points represents whether the given pair is selected as an outlier by the underlying algorithm. The meaning of this figure follows Fig.\ref{fig:diffvqa}.  } \label{fig:diffiqa}
	\end{center}
	\end{figure}


For simplicity, we randomly take
reference 1 as an illustrative example (other reference
videos exhibit similar results). We show the outlier detection results for both the LASSO+L2 algorithm and the LBI algorithm in Fig.\ref{fig:diffvqa}. Here we regard the top 5\% detection results as the source of outliers. In this figure, the shape of points represents whether the given pair is selected as an outlier by the underlying algorithm.  Each point in the figure represents a comparison pair (item1, item2), the $x$-axis gives the ID for item2 while the $y$-axis gives the ID with item1. The labels in the $x$- and $y$-axis are arranged according to the score order from the L2 method (ascendingly). The color of points represents the strength of outliers. We adopt a color gradient from red to blue to represent the top 5\%, 10\%, 15\%, 20\%, 25\% outliers and the other pairs.  It is interesting to see that the picked out outliers are mainly distributed in the lower-left corner of this plot. This is reasonable since such points  are the most suspicious results which bear significant deviation with the global ranking returned from the L2 method.   In addition, from the basic property of the regularization path, the earlier a pair is detected as an outlier (jumped out from the path), the larger the corresponding deviation is. This is visualized here with a darker and red color. In this sense, we see that both algorithms could spot the most suspicious pairs colored in dark red. Moreover, Fig.\ref{data5} further confirms this phenomenon. Here, all the top 5\% outliers show reverse results from the global ranking of L2 with arrows pointing from lower quality to higher quality videos. More importantly, it is easy to see that outliers returned by these two approaches are exactly the same. This shows that LBI could approximate the unbiased estimation nicely.

Tab.\ref{tab:data5-rank} shows the outcomes of all competitors. For the sake of convenience, we rearrange the IDs according to their resulting rank returned by the unbiased LASSO+L2 method. Since the LASSO+L2 method is a standard unbiased algorithm, a reasonable algorithm should return a consistent result with this method.  We can see that LBI succeeds in giving the most consistent result with LASSO+L2. The other algorithms, instead, exhibit a more significant difference with LASSO+L2.  Fig.\ref{fig:vqa_rank} provides another visualization by means of the Bump Chart. Paying a careful look, it is easy to see that the removal of the top 5\% outliers in both LASSO+L2 and LBI changes the orders of some competitive videos, such as V3 and V12. Instead, HLASSO remains the same result as the L2 algorithm. This indicates HLASSO is more biased than the other two and the effect of outliers is mainly within the highly competitive groups.

\begin{table*}[t]
		\caption{Comparison of different rankings on IQA data. Eight algorithms are compared here with the integer representing the ranking position and the number in parenthesis representing the global ranking score returned by the corresponding algorithm. See Tab.\ref{tab:data5-rank} for more explanations of the details. }\label{tab:IQA}
		\centering

		\begin{tabular}{lcccccccc}
			ID    & LASSO+L2 & L2    & XGB\_Linear & PCLASSO & iLTS  & iHT   & HLASSO & LBI \\
			\toprule
			1     & 1 ( 0.8876) & 1 ( 0.8001) & 1 ( 0.7967) & 1 ( 0.9835) & 1 ( 0.9059) & 1 ( 0.9031) & 1 ( 0.8144) & 1 ( 0.8851) \\
			6     & 2 ( 0.7034) & 2 ( 0.6003) & 2 ( 0.5974) & 2 ( 0.8384) & 2 ( 0.7711) & 2 ( 0.7402) & 2 ( 0.6143) & 2 ( 0.6977) \\
			9     & 3 ( 0.6048) & 3 ( 0.5362) & 3 ( 0.5327) & 3 ( 0.7654) & 3 ( 0.5600) & 3 ( 0.5650) & 3 ( 0.5484) & 3 ( 0.5929) \\
			12    & 4 ( 0.4886) & 4 ( 0.4722) & 4 ( 0.4699) & 4 ( 0.6303) & 4 ( 0.4706) & 4 ( 0.5126) & 4 ( 0.4752) & 4 ( 0.4895) \\
			2     & 5 ( 0.2859) & \textcolor{red}{6 ( 0.3044)} & \textcolor{red}{6 ( 0.3027)} & \textcolor{red}{6 ( 0.5458)} & 5 ( 0.3504) & \textcolor{red}{6 ( 0.3026)} & \textcolor{red}{6 ( 0.3105)} & 5 ( 0.3115) \\
			10    & 6 ( 0.2698) & \textcolor{red}{5 ( 0.3472)} & \textcolor{red}{5 ( 0.3451)} & \textcolor{red}{5 ( 0.5545)} & \textcolor{red}{7 ( 0.2590)} & \textcolor{red}{5 ( 0.3285)} & \textcolor{red}{5 ( 0.3368)} & 6 ( 0.2770) \\
			16    & 7 ( 0.2677) & 7 ( 0.2756) & 7 ( 0.2747) & 7 ( 0.5378) & \textcolor{red}{6 ( 0.2925)} & 7 ( 0.2640) & 7 ( 0.2757) & 7 ( 0.2680) \\
			7     & 8 ( 0.1398) & 8 ( 0.1403) & 8 ( 0.1396) & \textcolor{red}{9 ( 0.3002)} & 8 ( 0.1012) & 8 ( 0.0808) & 8 ( 0.1374) & 8 ( 0.1392) \\
			15    & 9 ( 0.0540) & 9 ( 0.0965) & 9 ( 0.0963) & \textcolor{red}{8 ( 0.4006)} & 9 (-0.0086) & 9 (-0.0075) & 9 ( 0.0865) & 9 ( 0.0418) \\
			11    & 10(-0.1815) & 10(-0.1609) & 10(-0.1597) & 10( 0.1308) & 10(-0.1889) & 10(-0.1649) & 10(-0.1563) & 10(-0.1739) \\
			8     & 11(-0.2813) & 11(-0.2541) & 11(-0.2525) & 11( 0.0217) & 11(-0.2527) & 11(-0.2753) & 11(-0.2620) & 11(-0.2803) \\
			13    & 12(-0.2927) & 12(-0.2964) & 12(-0.2946) & 12(-0.0121) & 12(-0.2699) & 12(-0.3124) & 12(-0.2958) & 12(-0.2929) \\
			3     & 13(-0.6246) & \textcolor{red}{14(-0.6315)} & \textcolor{red}{14(-0.6285)} & \textcolor{red}{14(-0.3239)} & \textcolor{red}{14(-0.7343)} & \textcolor{red}{14(-0.6506)} & 13(-0.6315) & 13(-0.6316) \\
			14    & 14(-0.6478) & \textcolor{red}{13(-0.6215)} & \textcolor{red}{13(-0.6181)} & \textcolor{red}{13(-0.2795)} & \textcolor{red}{13(-0.5932)} & \textcolor{red}{13(-0.5873)} & 14(-0.6361) & 14(-0.6799) \\
			4     & 15(-0.8098) & 15(-0.7822) & 15(-0.7791) & 15(-0.4022) & 15(-0.8090) & 15(-0.8124) & 15(-0.7889) & 15(-0.8102) \\
			5     & 16(-0.8639) & 16(-0.8262) & 16(-0.8229) & 16(-0.5608) & 16(-0.8539) & 16(-0.8863) & 16(-0.8287) & 16(-0.8339) \\
			\bottomrule
		\end{tabular}%
\end{table*}

\subsubsection{PC-IQA Dataset}
\begin{figure}[t]
	\begin{center}
		\includegraphics[width=\linewidth]{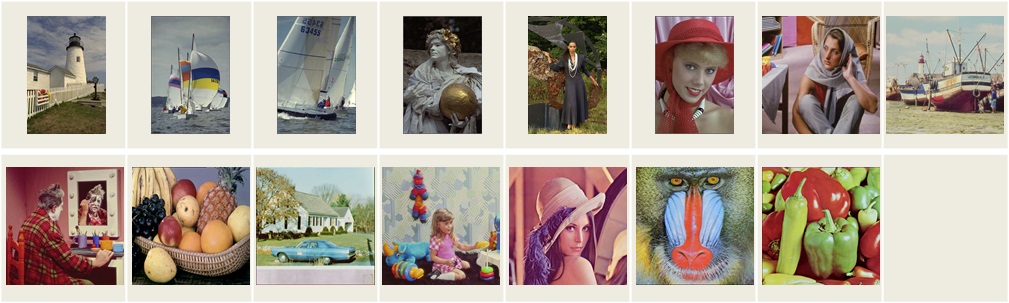}
		\caption{Reference images in LIVE and IVC databases. (The first six are from LIVE and the remaining nine are from IVC. Row 1: ID=1-8; Row 2: ID=9-15.)} \label{dataimages}
	\end{center}
\end{figure}

In the second dataset, we test the detection ability on incomplete and imbalanced data over the Image Quality Assessment dataset PC-IQA. This dataset contains 15 reference images and 15 distorted versions of each reference, for a total of 240 images which come from two publicly available datasets, LIVE
\citep{LIVE} and IVC \citep{IVC}, as shown in Fig.\ref{dataimages}. Totally, 186 observers,
each of whom performs a varied number of comparisons via Internet, provide
$23,097$ paired comparisons for crowdsourced subjective image quality assessment.

\begin{figure}[t]
	\begin{center}
		\includegraphics[width=0.9\columnwidth]{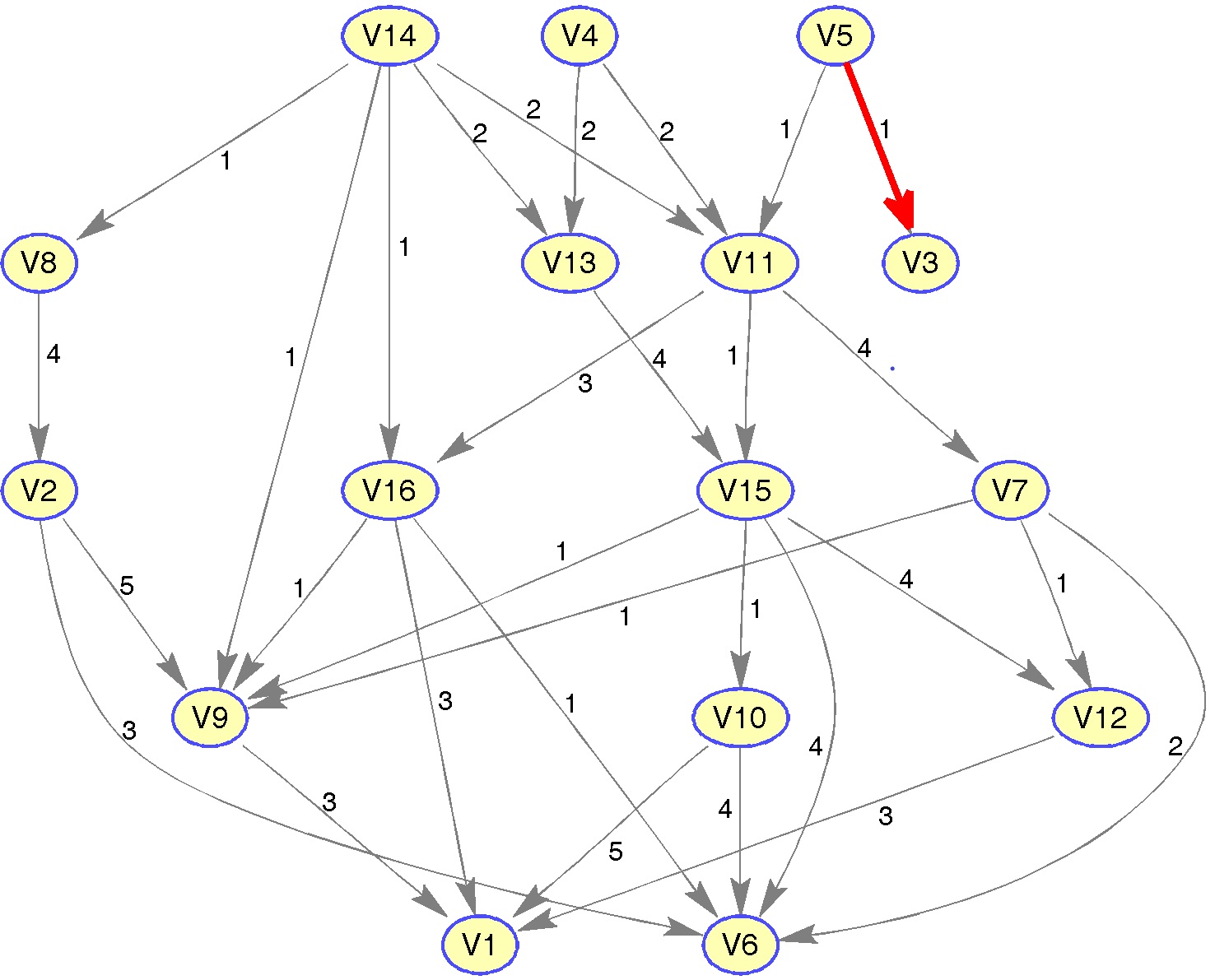}
		\caption{Top 5\% outliers for PC-IQA dataset. The red arrow means that LASSO+L2 treated the result as an outlier, while LBI believes that it is a normal comparison. } \label{data10}
	\end{center}
\end{figure}

Fig.\ref{fig:diffiqa} and Tab.\ref{tab:IQA} show the experimental results on a randomly selected reference (image 10 in Fig.\ref{dataimages}). Similar as the first dataset, the points located at the lower-left corner of Fig.\ref{fig:diffiqa} are supposed to be the suspicious results.  Similar observations as PC-VQA dataset can be made in this sense and we note that outliers distributed on this dataset are much sparser than PC-VQA, shown by many zeros in the lower left corner of Fig.\ref{fig:diffiqa}. Again, from Fig.\ref{data10}, we see these outliers provide the reversed preference results from the predicted global ranking from L2. Moreover, both Fig.\ref{fig:diffiqa} and Fig.\ref{data10} suggest that LBI and LASSO+L2 only disagree with each other at one pair, i.e., the comparison of $V_3$ \textit{vs.} $V_5$. A closer look at the Fig.\ref{fig:diffiqa} shows that this pair is located at border between top 5\% and top 10\% outliers, which is difficult to be detected.

\begin{figure}

		\begin{center}
			\includegraphics[width=\linewidth]{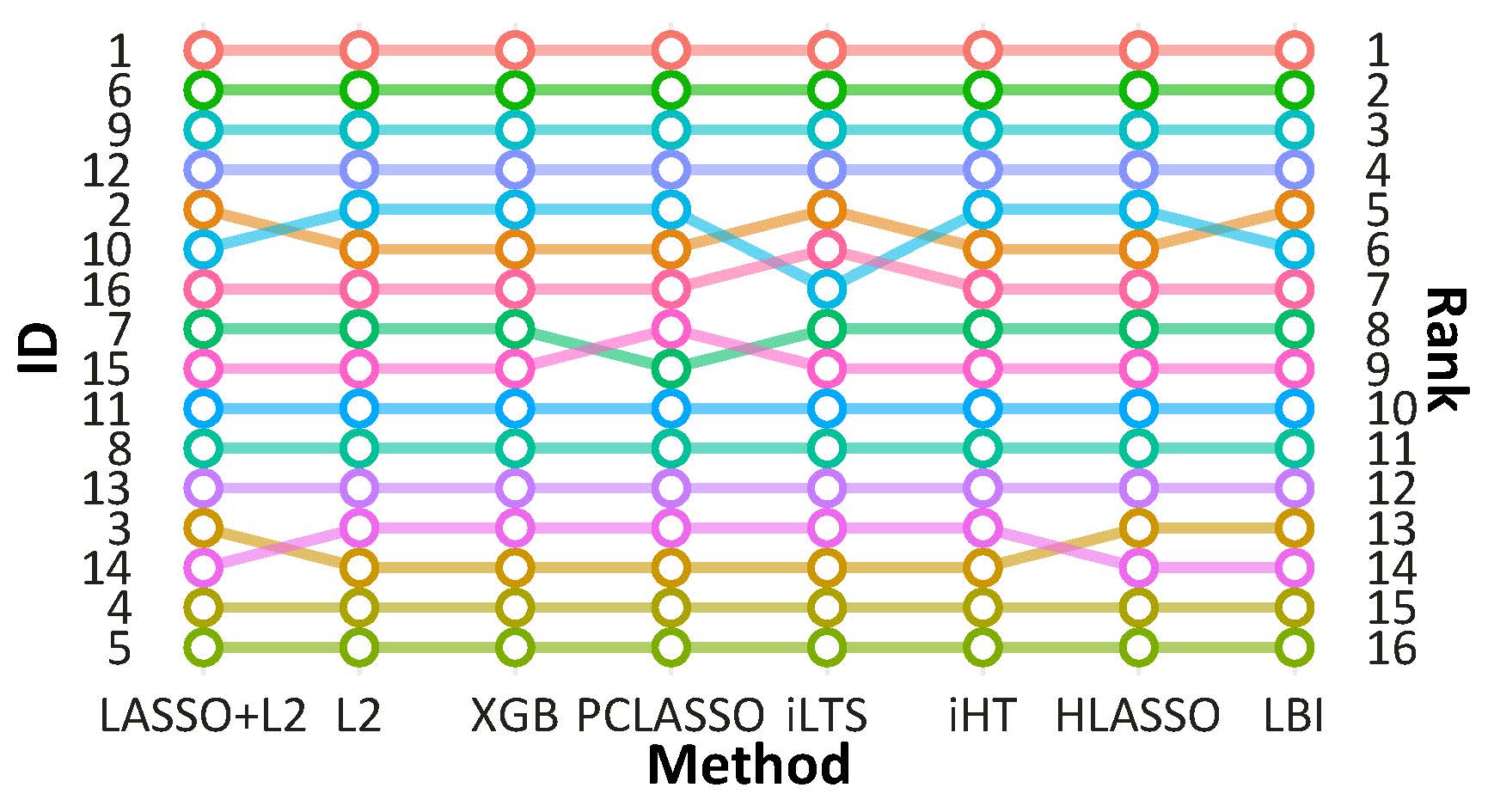}
			\caption{Visualization of the Rank on PC-IQA dataset with bump chart. See the caption of Fig.\ref{fig:vqa_rank} for more details.} \label{fig:iqa_rank}
		\end{center}
\end{figure}

Taking a step further, the outcomes of  all the competitors are shown in Tab.\ref{tab:IQA}. We can see that LBI succeeds in giving the most consistent result. The others, instead, exhibit a more significant difference with LASSO+L2. See Fig.\ref{fig:iqa_rank} for another visualization by means of the Bump Chart.

\begin{figure}
	\begin{center}
		\includegraphics[width=0.9\linewidth]{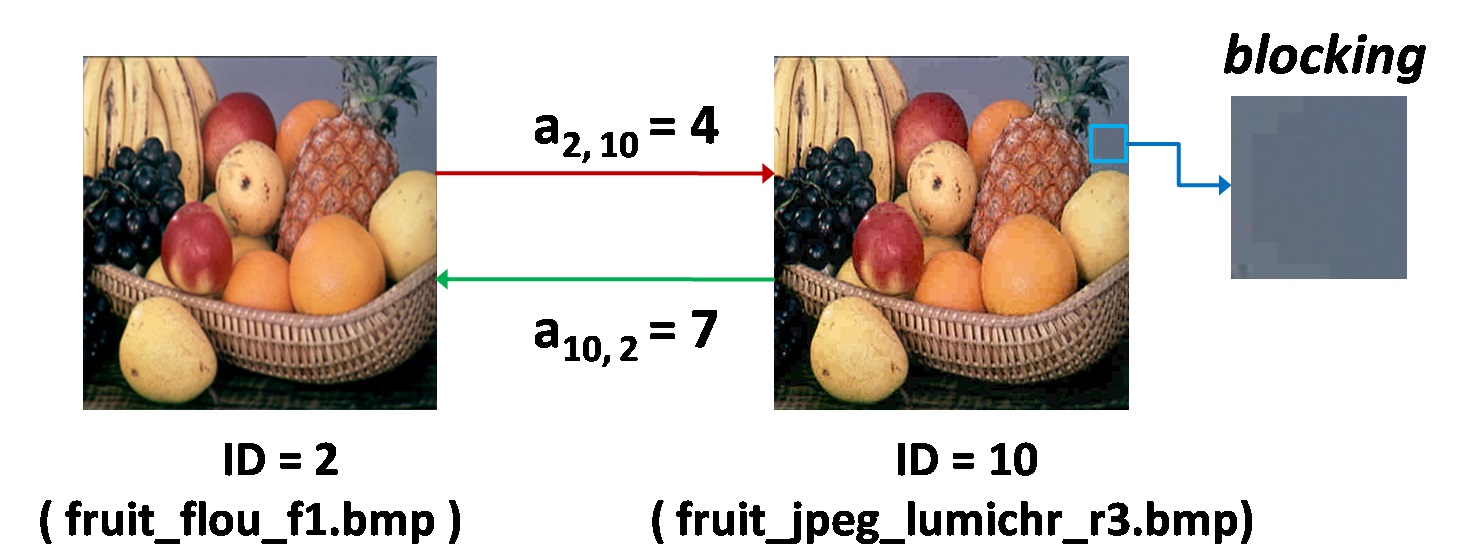}
		\caption{Dissimilar judgments due to multi-criteria in paired comparisons among users. The image is indistinguishable due to its small size, so image names in IVC \citep{IVC} are printed here.} \label{ivc2}
	\end{center}
\end{figure}

\begin{table*}[tp]
	\centering
	\caption{\label{tab:rank_age}Comparison of different rankings on Human age data. Eight algorithms are compared here with the integer representing the ranking position and the number in parenthesis representing the global ranking score returned by the corresponding algorithm. See Tab.\ref{tab:data5-rank} for more explanations of the details. }
	\begin{tabular}{lcccccccc}
		ID   & LASSO+L2 & L2    & XGB\_Linear & PCLASSO & iLTS  & iHT   & HLASSO & LBI \\
		\toprule
		28    & 1 ( 0.9192) & 1 ( 0.7780) & 1 ( 0.7772) & 1 ( 1.0021) & 1 ( 0.9559) & 1 ( 0.9192) & 1 ( 0.8173) & 1 ( 0.9177) \\
		3     & 2 ( 0.8310) & 2 ( 0.6661) & 2 ( 0.6654) & 2 ( 0.9681) & 2 ( 0.8585) & 2 ( 0.8311) & 2 ( 0.7050) & 2 ( 0.8268) \\
		14    & 3 ( 0.7350) & 3 ( 0.5653) & 3 ( 0.5647) & \textcolor{red}{4 ( 0.8160)} & 3 ( 0.7508) & 3 ( 0.7350) & 3 ( 0.5980) & 3 ( 0.7074) \\
		29    & 4 ( 0.5268) & 4 ( 0.4482) & 4 ( 0.4477) & \textcolor{red}{3 ( 0.8439)} & \textcolor{red}{5 ( 0.6191)} & 4 ( 0.5268) & 4 ( 0.4669) & 4 ( 0.5233) \\
		11    & 5 ( 0.5128) & \textcolor{red}{6 ( 0.4059)} & \textcolor{red}{6 ( 0.4055)} & 5 ( 0.7102) & \textcolor{red}{4 ( 0.6214)} & 5 ( 0.5127) & 5 ( 0.4304) & 5 ( 0.5103) \\
		21    & 6 ( 0.4800) & \textcolor{red}{5 ( 0.4087)} & \textcolor{red}{5 ( 0.4083)} & \textcolor{red}{8 ( 0.6682)} & 6 ( 0.5599) & 6 ( 0.4799) & 6 ( 0.4257) & 6 ( 0.4786) \\
		5     & 7 ( 0.4594) & \textcolor{red}{8 ( 0.3634)} & \textcolor{red}{8 ( 0.3631)} & \textcolor{red}{6 ( 0.7037)} & \textcolor{red}{8 ( 0.5464)} & 7 ( 0.4593) & \textcolor{red}{8 ( 0.3824)} & 7 ( 0.4554) \\
		7     & 8 ( 0.4535) & \textcolor{red}{7 ( 0.3873)} & \textcolor{red}{7 ( 0.3869)} & \textcolor{red}{7 ( 0.7015)} & \textcolor{red}{7 ( 0.5550)} & 8 ( 0.4535) & \textcolor{red}{7 ( 0.4028)} & 8 ( 0.4520) \\
		27    & 9 ( 0.4342) & 9 ( 0.3582) & 9 ( 0.3578) & 9 ( 0.6170) & 9 ( 0.4979) & 9 ( 0.4341) & 9 ( 0.3755) & 9 ( 0.4318) \\
		24    & 10( 0.2792) & 10( 0.2064) & 10( 0.2062) & 10( 0.5344) & 10( 0.2950) & 10( 0.2793) & 10( 0.2217) & 10( 0.2677) \\
		6     & 11( 0.1365) & 11( 0.0932) & 11( 0.0931) & 11( 0.5030) & 11( 0.1379) & 11( 0.1364) & 11( 0.1002) & 11( 0.1290) \\
		17    & 12( 0.1205) & \textcolor{red}{14( 0.0872)} & \textcolor{red}{14( 0.0871)} & \textcolor{red}{13( 0.4482)} & \textcolor{red}{15( 0.0687)} & \textcolor{red}{14( 0.1099)} & \textcolor{red}{14( 0.0923)} & \textcolor{red}{14( 0.1099)} \\
		4     & 13( 0.1202) & \textcolor{red}{12( 0.0914)} & \textcolor{red}{12( 0.0913)} & \textcolor{red}{12( 0.4562)} & 13( 0.1099) & \textcolor{red}{12( 0.1202)} & \textcolor{red}{12( 0.0962)} & \textcolor{red}{12( 0.1158)} \\
		22    & 14( 0.1099) & \textcolor{red}{13( 0.0896)} & \textcolor{red}{13( 0.0895)} & \textcolor{red}{16( 0.4277)} & \textcolor{red}{12( 0.1147)} & \textcolor{red}{13( 0.1133)} & \textcolor{red}{13( 0.0944)} & \textcolor{red}{13( 0.1125)} \\
		20    & 15( 0.1070) & 15( 0.0816) & 15( 0.0815) & 15( 0.4283) & \textcolor{red}{14( 0.0808)} & 15( 0.1070) & 15( 0.0870) & 15( 0.1050) \\
		23    & 16( 0.0305) & 16( 0.0208) & 16( 0.0208) & \textcolor{red}{18( 0.3405)} & 16( 0.0174) & 16( 0.0304) & 16( 0.0233) & 16( 0.0310) \\
		8     & 17( 0.0185) & 17( 0.0086) & 17( 0.0085) & 17( 0.3654) & 17(-0.0093) & 17( 0.0187) & 17( 0.0109) & 17( 0.0196) \\
		30    & 18( 0.0003) & 18(-0.0025) & 18(-0.0025) & \textcolor{red}{14( 0.4373)} & 18(-0.0404) & 18( 0.0001) & 18(-0.0030) & 18( 0.0000) \\
		12    & 19(-0.0358) & 19(-0.0201) & 19(-0.0200) & \textcolor{red}{20( 0.2877)} & 19(-0.0684) & 19(-0.0358) & 19(-0.0226) & 19(-0.0324) \\
		13    & 20(-0.2527) & 20(-0.1961) & 20(-0.1959) & \textcolor{red}{21( 0.2459)} & 20(-0.3133) & 20(-0.2524) & 20(-0.2073) & 20(-0.2568) \\
		15    & 21(-0.2629) & 21(-0.2160) & 21(-0.2158) & \textcolor{red}{22( 0.1844)} & 21(-0.3170) & 21(-0.2658) & 21(-0.2248) & 21(-0.2645) \\
		25    & 22(-0.2751) & 22(-0.2166) & 22(-0.2164) & \textcolor{red}{19( 0.2924)} & 22(-0.3338) & 22(-0.2751) & 22(-0.2281) & 22(-0.2754) \\
		16    & 23(-0.3583) & 23(-0.2551) & 23(-0.2548) & 23( 0.1577) & 23(-0.4359) & 23(-0.3583) & 23(-0.2706) & 23(-0.3454) \\
		2     & 24(-0.4633) & 24(-0.3710) & 24(-0.3706) & \textcolor{red}{25( 0.0657)} & 24(-0.5184) & 24(-0.4565) & 24(-0.3881) & 24(-0.4407) \\
		9     & 25(-0.5618) & 25(-0.4158) & 25(-0.4154) & \textcolor{red}{24( 0.0670)} & 25(-0.6058) & 25(-0.5617) & 25(-0.4408) & 25(-0.5376) \\
		1     & 26(-0.7429) & 26(-0.6135) & 26(-0.6129) & 26(-0.0478) & 26(-0.7520) & 26(-0.7334) & 26(-0.6437) & 26(-0.7317) \\
		18    & 27(-0.7436) & 27(-0.6249) & 27(-0.6243) & 27(-0.1575) & 27(-0.7686) & 27(-0.7434) & 27(-0.6548) & 27(-0.7418) \\
		19    & 28(-0.7624) & 28(-0.6653) & 28(-0.6647) & \textcolor{red}{30(-0.2651)} & 28(-0.7826) & 28(-0.7688) & 28(-0.6907) & 28(-0.7631) \\
		10    & 29(-0.8741) & 29(-0.6969) & 29(-0.6962) & \textcolor{red}{28(-0.2153)} & 29(-0.8921) & 29(-0.8739) & 29(-0.7429) & 29(-0.8664) \\
		26    & 30(-0.9417) & 30(-0.7660) & 30(-0.7652) & \textcolor{red}{29(-0.2455)} & 30(-0.9517) & 30(-0.9416) & 30(-0.8126) & 30(-0.9380) \\
		\bottomrule
	\end{tabular}%
\end{table*}%

Now we provide a closer look at some tiny difference between the most results from HLASSO, LBI and LASSO+L2. Based on 5\% outliers detection, HLASSO, LBI $(50,1/25000)$, and LASSO+L2 all differ with L2 in that image ID = 3 ({\tt fruit\_{}flou\_{}f3.bmp} in IVC \citep{IVC}) is better than image ID = 14 ({\tt fruit\_{}lar\_{}r1.bmp}). Such a preference is in agreement with the pairwise majority voting of 9:6 votes. Therefore, the example shows that under sparse outliers L2 ranking may be less accurate. Moreover, LASSO+L2 and LBI further suggest that image ID = 2 should be better than image ID = 10, in contrast to L2 and HLASSO algorithms. Further inspection of the dataset confirms that such a suggestion is reasonable. Fig.\ref{ivc2} shows the two images (ID = 2 and ID = 10) from the IVC \citep{IVC} database.
There is a blurring effect in image ID = 2 and a blocking effect in the background of ID = 10.
In the dataset, 4 raters agree that the quality of ID = 2 is better than that from ID = 10. By contrast,
while 7 raters have the opposite opinion. Clearly both LASSO+L2 and LBI choose the preference of the minority,
based on aggregate behavior over population after removal of some outliers. Why does this happen?
In fact, when a participant compares the quality between ID = 2 and ID = 10, his/her preference depends on his/her
attention --- on the foreground or on the whole image.
A rater with foreground attention might be disturbed by the blurring effect, leading to $10\succ 2$.
On the other hand, a rater with holistic attention may notice the blocking effect in the background, leading to $2\succ 10$.
Which criterion might be dominant? To explore this question, we further collected cleaner data (i.e., 20 more persons provide careful judgments in controlled lab conditions),
among which a dominant percentage (80\%) agrees with $2\succ 10$. This is  consistent with the LASSO+L2/LBI prediction after removal of outliers.
This suggests that most observers assess the quality of an image from a global point of view instead of a local one.
Another less stable way is to select a subset of clean data without outliers for the validation procedure.
On one hand, such a result suggests that for those highly competitive or confusing alternative pairs,
a large number of samples are expected to find a good ranking in majority voting.
On the other hand, by exploiting intermediate comparisons of good confidence with other alternatives,
it is possible to achieve a reliable global ranking with a much smaller number of samples, such as what both LASSO+L2 and LBI
do here, which suggests that our proposed methods bear smaller sample complexity.

\subsubsection{Human Age Dataset}

\begin{figure}
	\begin{center}
		\includegraphics[width=\linewidth]{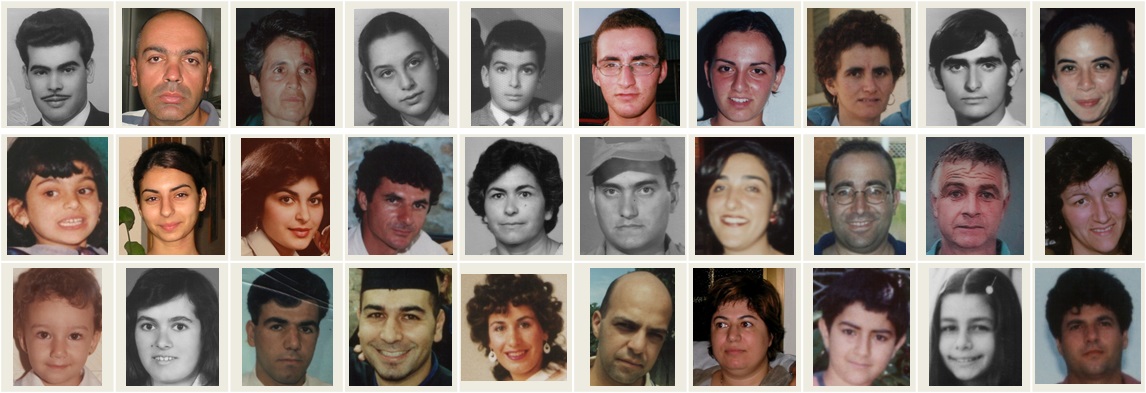}
		\caption{Images in Human age dataset.} \label{agedataset}
	\end{center}
\end{figure}

\begin{figure}
	\begin{center}
		\includegraphics[width=\linewidth]{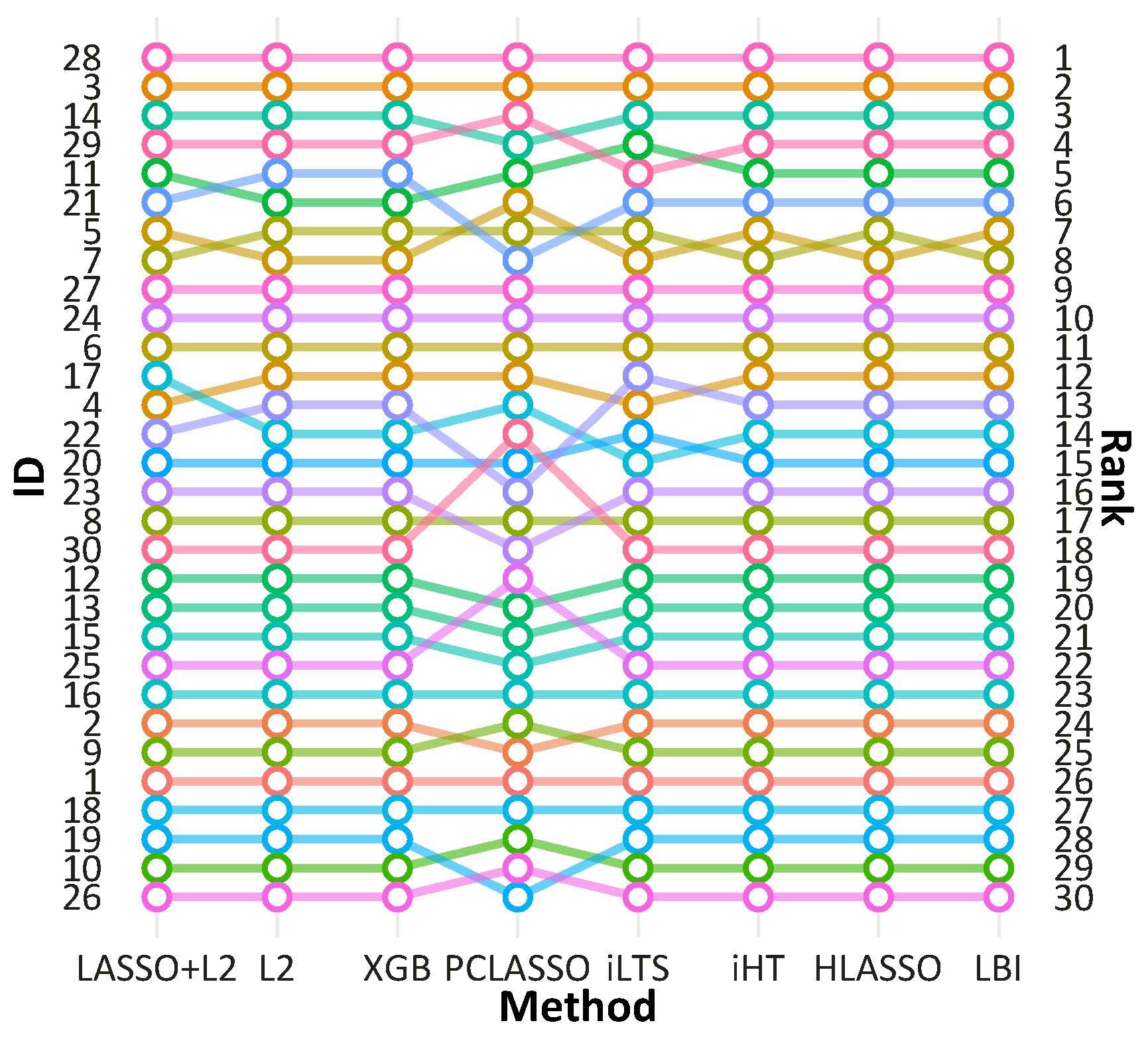}
		\caption{\label{fig:age_rank}Visualization of the Rank on Human Age dataset with bump chart. See the caption of Fig.\ref{fig:vqa_rank} for more details.} \label{age_rank}
	\end{center}
\end{figure}

Now we provide our experimental analysis about continuous ranking in the third dataset called Human Age. In this dataset, 30 images from human age dataset FG-NET\footnote{http://www.fgnet.rsunit.com/} are annotated by a group of volunteer users on \href{http://www.chinacrowds.com/}{ChinaCrowds} platform, as is illustrated in Fig.\ref{agedataset}. The annotator is presented with two images and
given a binary choice of which one is older. We obtain 14,011 pairwise comparisons from 94 annotators.

Firstly Tab.\ref{tab:rank_age} and Fig.\ref{fig:age_rank} provide  fine-grained comparisons of the ranking results from all the competitors.  In this sense, we can see that LBI succeeds in giving the most consistent result. The other algorithms, instead, exhibit a more significant difference with LASSO+L2. Without a robust mechanism to resist the outliers existed in the dataset, it turns out that L2, XGB\_Linear, and PCLASSO provide the most inconsistent results. iLTS, iHT and HLASSO, from another perspective, provide less consistent results due to their ability to detect outliers. However, since none of them could approximate an unbiased solution like LBI, their difference with LASSO+L2 is still more apparent compared with the proposed LBI. See Fig.\ref{fig:age_rank} for another visualization through the Bump Chart.

\begin{figure}
\begin{center}
	\includegraphics[width=\linewidth]{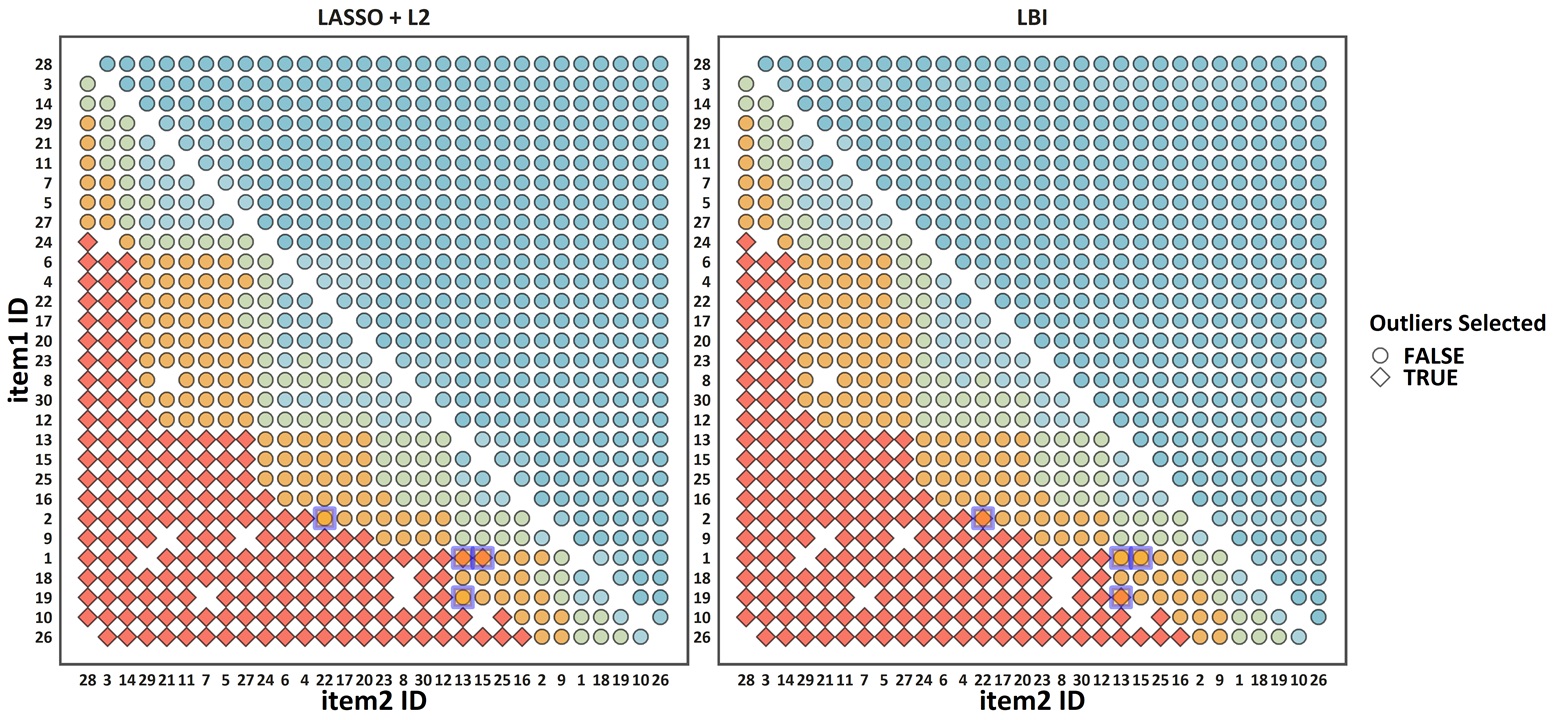}
	\caption{Visualization of the outliers in the Human Age dataset. The shape of points represents whether the given pair is selected as an outlier by the underlying algorithm. The meaning of this figure follows Fig.\ref{fig:diffvqa}.      } \label{fig:diffage}
\end{center}
\end{figure}

Next, we are ready to have a closer investigation into the outliers selected by LBI and LASSO+L2, which is presented in Fig.\ref{fig:diffage}, where we select the top 5\% pairs which jump out in the earliest time from the regularization path as the outliers. Here one can clearly find that our proposed algorithm LBI shares a similar result with LASSO+L2, where only four times they disagree with each other.  Note that the x- and y-axis are the IDs rearranged by the rank (descending) produced by L2. The pair $(x,y)$ with $x \succ y$ (located at the lower-left corner) provides a conflicted result with the majority of the votes. In this sense, we see that both algorithms could successfully figure out the outliers.


\subsubsection{Face Verification Dataset} \label{sec:LFW}
\begin{figure*}
	\begin{center}
		\includegraphics[width=\linewidth]{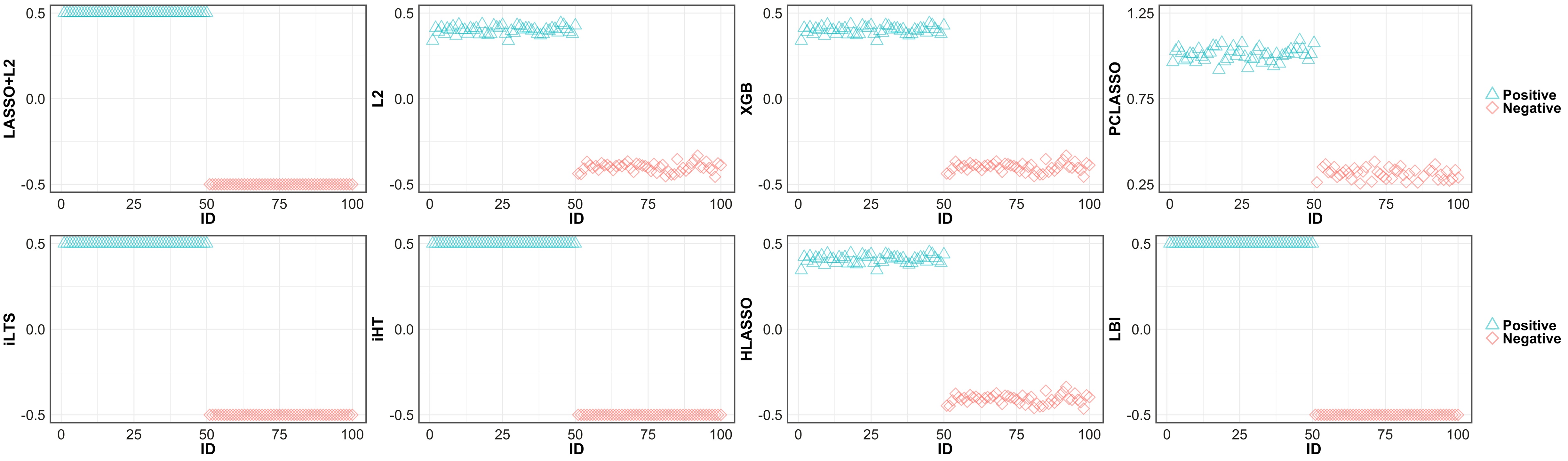}
		\caption{Bipartite ranking results on FGLFW dataset.} \label{fig:biface}
	\end{center}
\end{figure*}

\begin{figure*}
	\begin{center}
		\includegraphics[width=\linewidth]{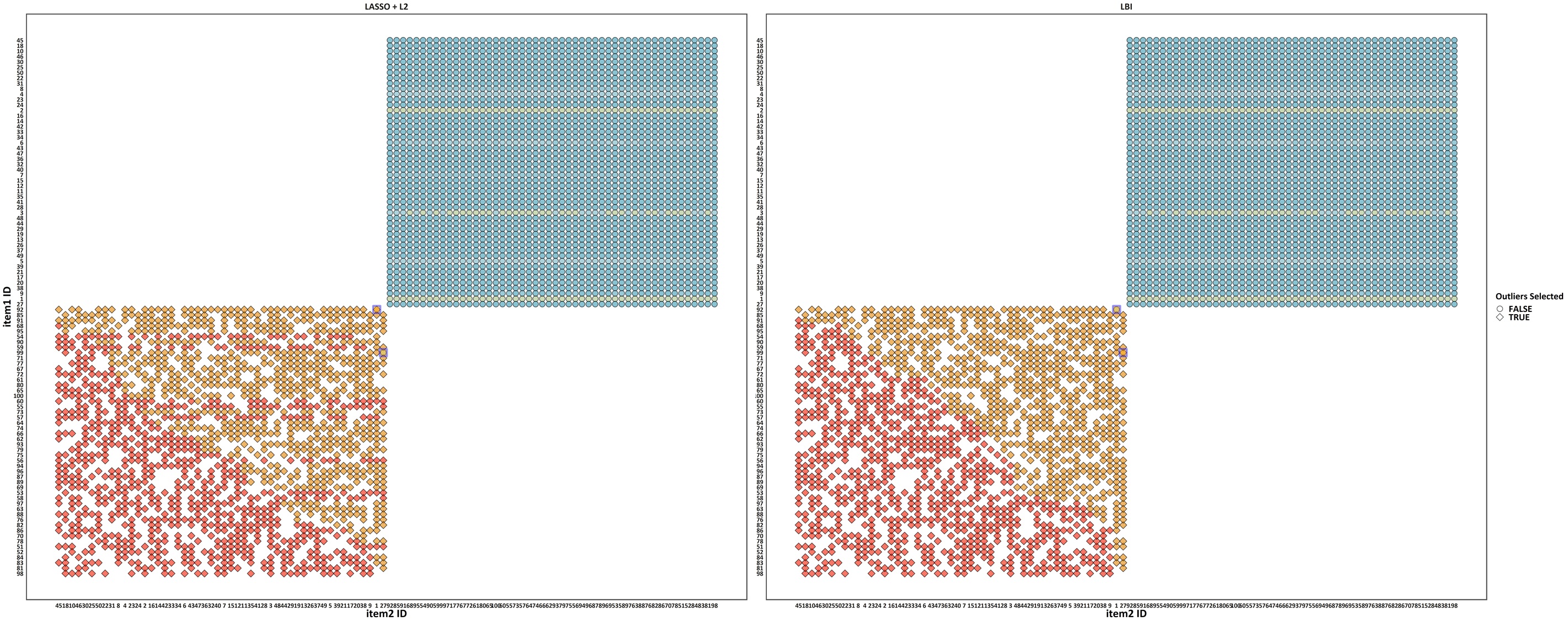}
		\caption{Visualization of the outliers in the FGLFW dataset. The shape of points represents whether the given pair is selected as an outlier by the underlying algorithm. The meaning of this figure follows Fig.\ref{fig:diffvqa}.     } \label{fig:difflfw}
	\end{center}
\end{figure*}

In the previous experiments, we demonstrate the effectiveness of our proposed method for continuous ranking problems.
Next, we shift our focus to a specific discrete ranking problem called bipartite ranking. Given a set of positive instances and negative instances, the goal of bipartite ranking is to render all positive instances are higher rank than the negative instances. This is crucial for applications like image retrieval where the top-ranked instances should be relevant samples.

We start with a face comparison dataset: Fine-grained LFW (FGLFW) \footnote{\url{http://www.whdeng.cn/FGLFW/FGLFW.html}}, which is a renovation of the well-known Label Faces in the Wild (LFW) dataset \footnote{\url{http://vis-www.cs.umass.edu/lfw/}}  containing more complicated human face comparisons with finer-grained differences. In FGLFW, there are 10 sets of fine-grained human face comparisons, each of which containing 300 matched pairs and 300 mismatched pairs. For this dataset, we regard each human face pair as an instance and perform a similarity learning task where we require the matched pairs having a higher score than the mismatched ones.  For the sake of convenience, we generate a noisy crowdsourced dataset based on the first set of pairs in this dataset. To do this, we first randomly select 50 matched and mismatched pairs, correspondingly. To obtain a set of crowdsourced pairwise comparisons, we then pick up all the pairs formed by a matched pair and a mismatched pair. For each of the pair $(i,j)$ with $i$ being a matched face pair and $j$ being a mismatched one, we randomly generate $n_{i,j} \sim \mathcal{U}(5,15)$ times of personalized comparisons with annotation $i \succ j$ and we randomly flip the resulting annotations to ensure that there are 10\% of noisy annotations conflicted with the ground-truth result. In the forthcoming content, we present our experimental results in two dimensions.

First of all, Fig.\ref{fig:biface} provides the fine-grained comparison results for different algorithms. Here the red points represent the truly matched pairs and the blue points represent the mismatched ones. Recall that all the possible pairs between negative samples and positives samples are observed in our dataset, a reasonable ranker should then satisfy the following two properties: (1) render the same rank for all pairs coming from the same cluster (2) differentiate the matched cluster and mismatched cluster with a clear score gap. From the figure, we see that only LBI, iLTS, iHT, and LASSO+L2 satisfy such properties. The other algorithms, instead, show different degrees of score perturbations. The results again suggest that our proposed algorithm could provide consistent results with the unbiased LASSO+L2 method.

\begin{figure*}
		\begin{center}
			\includegraphics[width=\linewidth]{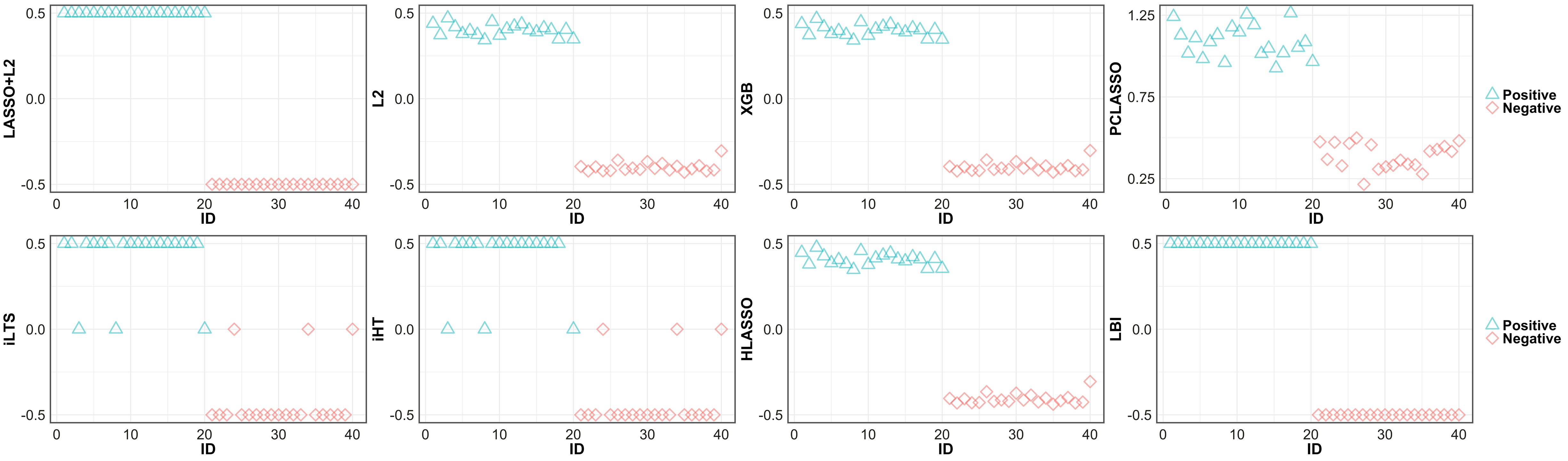}
			\caption{Bipartite ranking results on shoes dataset.} \label{fig:bishoes}
		\end{center}
\end{figure*}

From another dimension, Fig.\ref{fig:difflfw} gives us a fine-grained comparison between the outlier detection result generated by LASSO+L2 (LBI resp.).  Note that all the matched pair IDs are at most 50, while the mismatched IDs are all greater than 50. This means that all the pairs located at the lower-left corner in the plots are essentially outliers. Correspondingly, both LBI and LASSO+L2 could roughly pick out the majority of the outliers following this spirit. The blue boxed points show that they only disagree with each other at two pairs,  all of which lie in the border across normal and noisy observations.




\begin{figure}
		\begin{center}
			\includegraphics[width=\linewidth]{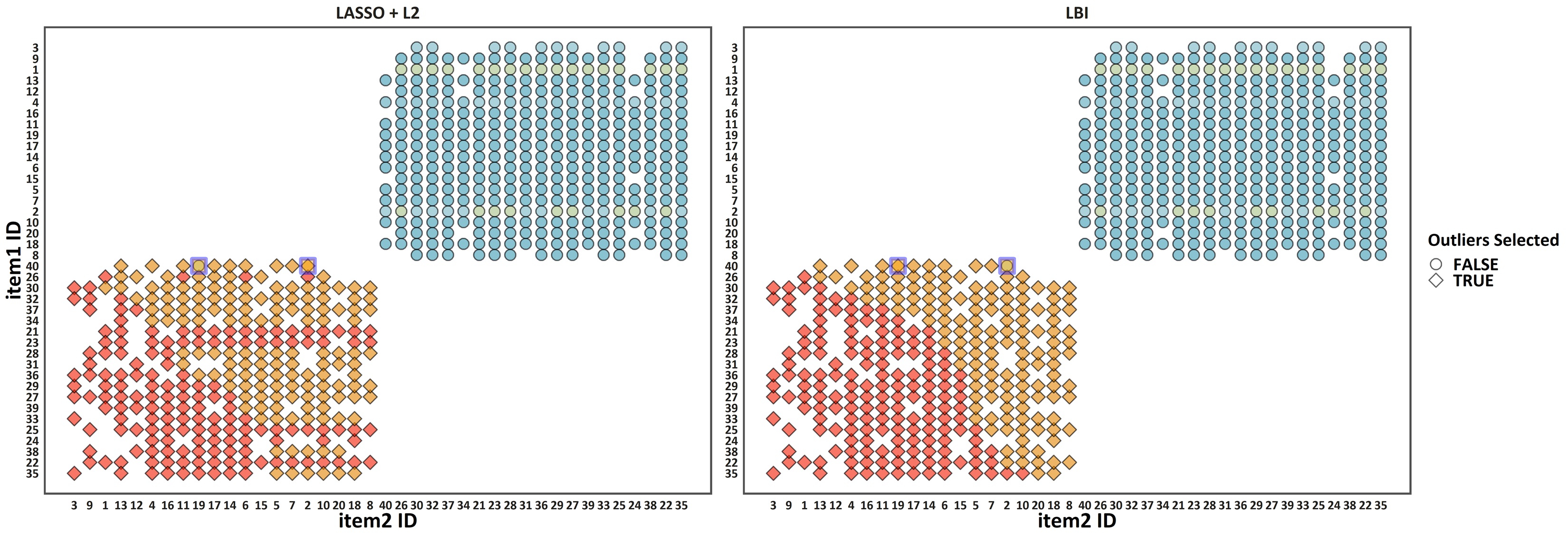}
			\caption{Visualization of the outliers in the Shoes dataset. The shape of points represents whether the given pair is selected as an outlier by the underlying algorithm. The meaning of this figure follows Fig.\ref{fig:diffvqa}.     } \label{fig:diffshoes}
		\end{center}
\end{figure}

\subsubsection{Shoes Dataset} \label{sec:shoes}

In this subsection, we proceed to explore the performance of our proposed algorithms for fine-grained retrieval applications. The Shoes dataset is collected from \cite{kovashka2015discovering} which contains 14,658 online shopping images. In this dataset,  a number of attributes are annotated by users with a wide spectrum of interests and backgrounds.  Similar to the FGLFW dataset, this dataset uses binary annotations rather than pairwise annotations (1 for Yes, -1 for No). Our goal here is to perform bipartite ranking to separate apart the positive instances and negative instances so that the top retrieved items are all true positive instances. Next, we narrow our attention to a specific attribute  Brown and generate a pairwise crowdsourcing dataset based on the original one to perform robust crowdsourced ranking. First, we conduct a majority vote for each of the labeled images in the original dataset as the reference label.  Then, 20 positive (negative resp.) instances are sampled randomly based on the resulting reference label. To further form the pairwise comparison, we then pick up all the possible pairs formed by a positive instance and a negative instance. For each of the pair $(i,j)$ with $i$ being a positive instance and $j$ being a negative one, we randomly generate $n_{i,j} \sim \mathcal{U}(5,100)$ times of personalized comparisons with annotation $i \succ j$ and we randomly flip the resulting annotations to ensure that there are 10\% of noisy annotations conflicted with the majority voting results.

Now we are ready to present the experimental results under the above-mentioned setting. First of all, Fig.\ref{fig:bishoes} provides the fine-grained comparison results for different algorithms. Here the red points represent the true positive instances and the blue points represent the true negative samples. Recall that all the possible pairs between negative samples and positives samples are observed in our dataset, a reasonable ranker should then satisfy the following two properties: (1) render all the positive instances (negative instance resp.) the same rank (score) and (2) differentiate the positive and negative instances with a clear score gap. From the figure, we see that only LBI and the unbiased LASSO+L2 satisfy such properties. The others, instead, show different degrees of score perturbations. As shown in the results for iLTS and iHT, such perturbations sometimes make it impossible to completely differentiate the relevant instances and irrelevant instances. The results again suggest that our proposed algorithm could provide consistent results with the unbiased LASSO+L2 method.

From another perspective, Fig.\ref{fig:diffshoes} gives us a fine-grained comparison between the outlier detection result generated by LASSO+L2 (LBI resp.) Note that all the positive instance IDs are at most 20, while the negative instance IDs are all greater than 20. This means that all the pairs located at the lower-left corner in the plots are essential outliers. Correspondingly, both LBI and LASSO+L2 could roughly pick out the majority of the outliers following this spirit. The blue boxed points show that they only disagree with each other at two pairs,  all of which lie in the border across normal and noisy observations.

\section{Conclusions}\label{sec:conclusions}
In this paper, we propose a framework of robust ranking for visual property evaluation, based on the principle of Hodge Decomposition. Theoretically, we find that outlier detection can be formulated as a sparse approximations of cyclic ranking projection in Hodge decomposition, which can be solved using Huber LASSO. Moreover, we propose another simple and effective method with Linearized Bregman Iterations to reduce the bias and automatically choose model parameters. Statistical consistency theory is established for both cases. Experimental studies are conducted with both simulated examples and real-world crowdsourcing data for visual property evaluation. Our results suggest the efficacy of our proposed methods.

\begin{acknowledgements}
This work was supported in part by the National Key R\&D Program of China under Grant No. 2018AAA0102003, in part by National Natural Science Foundation of China: 61861166002, U1736219, 61976202, U1803264, 61620106009, 61931008 and 61836002, in part by Youth Innovation Promotion Association CAS, and in part by the Strategic Priority Research Program of Chinese Academy of Sciences, Grant No. XDB28000000. The research of Yuan Yao was supported in part by Hong Kong Research Grant Council (HKRGC) grant 16303817, ITF UIM/390, as well as awards from Tencent AI Lab, Si Family Foundation, and Microsoft Research-Asia. 
\end{acknowledgements}

%
%

\bibliographystyle{spbasic}      
\bibliography{sigproc}   

\clearpage
\setcounter{page}{1}
\section*{Supplementary Material}
\section{Proof}

\begin{proof}[\textbf{Proof of Theorem \ref{thm:LASSO}}]
	The proof essentially follows the treatment in \citep{Wainwright09}. First of all, the equation \eqref{eq:HLASSO22} can been rewriten as
	\begin{equation}
	\min_{\gamma} \frac{1}{2} \| \tilde{y} - \Psi \gamma \|_2^2 + \lambda \|\gamma\|_{1}.
	\end{equation}
	where $\tilde{y} = \Psi_S \ga^\ast_S +  \Psi \varepsilon$, and $\varepsilon$ is sub-Gaussian with variance proxy $\sigma^2$.

	The duality implies that $\hat{\ga}$ is the unique minimizer if the following holds for some $\hat{v}\in \R^m$
	\[ -\Psi^T ( \tilde{y} - \Psi  \hat{\ga} ) + \la \hat{v} =0\]
	where for $i\in \hat{S}:=\supp(\hat{\ga})$, $\hat{v}_i = \sign(\hat{\ga}_i)$ and otherwise $\hat{v}_i \in (-1,1)$.

	To ensure that $\hat{S}\subseteq S=\supp(\ga^\ast)$, we just need for $j\in S^c$, $|\hat{v}_j|<1$, i.e.
	\begin{equation}  \label{eq:dualbound}
	\| \Psi_{S^c}^T (\tilde{y} - \Psi_{S}  \hat{\ga}_S) \|_\infty < \la,
	\end{equation}
	which ensures that $\hat{\ga}_j = 0$ for $j\in S^c$.
	On the other hand, for $i\in S$, we obtain the following equation
	\[ -\Psi_S^T (\tilde{y} - \Psi_S \hat{\ga}_S) + \la \hat{v}_S = 0, \]
	which leads to the following solution if $\Psi_S^T \Psi_S$ is invertible
	\begin{equation} \label{eq:dualeq}
	\hat{\ga}_S = \ga^\ast_S +  \delta_S, \ \ \ \delta_S:=(\Psi_S^T \Psi_S)^{-1} \left[ \Psi_S^T \Psi \varepsilon - \la\hat{v}_S\right] .
	\end{equation}
	Plugging (\ref{eq:dualeq}) into (\ref{eq:dualbound}) we have
	\[\| \Psi_{S^c}^T \Psi \tilde{\varepsilon} - \Psi_{S^c}^T\Psi_{S} (\Psi_S^T \Psi_S)^{-1} \left[ \Psi_S^T \Psi \tilde{\varepsilon} - \la \hat{v}_S \right]\|_\infty < \la \]
	or equivalently
	\[\| \la \Psi_{S^c}^T\Psi_{S} (\Psi_S^T \Psi_S)^{-1} \hat{v}_S  +\Psi_{S^c}^T( I -P_S) \Psi \tilde{\varepsilon} \|_\infty < \la .\]
	where $P_S=\Psi_{S} (\Psi_S^T \Psi_S)^{-1}\Psi_S^T$. Note that
	\begin{eqnarray*}
		&&\| \la \Psi_{S^c}^T\Psi_{S} (\Psi_S^T \Psi_S)^{-1} \hat{v}_S\|_\infty \\
		&\le& \la \|\Psi_{S^c}^T\Psi_{S} (\Psi_S^T \Psi_S)^{-1}\|_\infty\| \hat{v}_S\|_\infty\\
		&\le& \la(1-\eta).
	\end{eqnarray*}
	So it suffices to prove $\|\Psi_{S^c}^T( I -P_S) \Psi \tilde{\varepsilon}\|_\infty < \la\eta$.

	Now consider $Z_c = \Psi_{S^c}^T( I -P_S) (\Psi\varepsilon )$, for each $j\in S^c$ whose variance can be bounded by
	\begin{eqnarray*}
		\var(Z_c^{(j)}) & = &  \sigma^2 \Psi_{j}^T( I -P_S)^2 \Psi_{j}  \\
		& \leq & \sigma^2\mu_{\Psi}
	\end{eqnarray*}
	Hence
	\begin{eqnarray*}
		&&\Prob(\| \Psi_{S^c}^T( I -P_S) \Psi\varepsilon \|_\infty \geq \la\eta ) \\
		&\le& 2(l-s)\exp\left ( -\frac{\la^2\eta^2}{2 \sigma^2 \mu_{\Psi}  } \right)\\
		&=& 2 (l-s)m^{-\frac{2\la^2}{\overline{\la}^2}}\\
		&\leq&\frac{2(l-s)}{m^2},~~~ \la \ge \overline{\la}
	\end{eqnarray*}
	So, there is no false positive with probability at least $1-2/m$ if $\la \ge \overline{\la}$.

	To ensure the sign consistency, let $Z= (\Psi_S^T \Psi_S)^{-1} \Psi_S^T \Psi \varepsilon$, and for each $i\in S$,
	\begin{eqnarray*}
		\var(Z) & = & \sigma^2(\Psi_S^T \Psi_S)^{-1} \Psi_S^T \Psi\Psi^T \Psi_S (\Psi_S^T \Psi_S)^{-1}  \\
		& = & \sigma^2(\Psi_S^T \Psi_S)^{-1} \leq \frac{\sigma^2}{C_{min}} I.
	\end{eqnarray*}
	Then
	\[ \Prob(\|(\Psi_S^T \Psi_S)^{-1} \Psi_S^T \Psi \varepsilon\|_\infty \geq t ) \leq 2 |S|  \exp\left ( -\frac{t^2  C_{min}}{2\sigma^2} \right).\]
	Choose
	\[ t=\frac{\lambda \eta}{\sqrt{C_{min} \mu_{\Psi}}  }, \]
	then there holds
	\begin{eqnarray*}
		&&\Prob\left(\|Z\|_\infty \ge \frac{\la \eta}{ \sqrt{C_{min} \mu_{\Psi}} }\right) \\
		&\le& 2s\exp\left ( -\frac{\la^2\eta^2 }{2 \sigma^2  \mu_{\Psi}  } \right)\\
		&\le& \frac{2s}{m^2}.
	\end{eqnarray*}
	So, at least with probability $1-\frac{2}{m}$,
	\begin{eqnarray*}
		\|\hat{\ga}_S - \ga^\ast_S\|_\infty& = & \|\delta_S\|_\infty\\
		&\le& \|Z\|_\infty + \la \|(\Psi_S^T \Psi_S)^{-1}\hat{v}_S\|\\
		&\le& \frac{\la \eta}{ \sqrt{C_{min}} \mu_{\Psi} }+ \la \|(\Psi_S^T \Psi_S)^{-1}\|_\infty\\
		& = & c_0\la < \ga_{min},
	\end{eqnarray*}
	which means $\sign(\hat{\ga}) = \sign(\ga^\ast)$.

	Then the $l_2$-bound is straightly derived from
	\[\|\hat{\ga}_S - \ga^\ast_S\|_2 \le \sqrt{s}\|\hat{\ga}_S - \ga^\ast_S\|_\infty = c_0\sqrt{s}\la.\]
	This ends the proof.
\end{proof}

\bigskip

\begin{proof}[\textbf{Proof of Theorem \ref{thm:LB}}] The proof essentially extends the Theorem 4.2 in \citep{osher2014} from the Gaussian noise case to the sub-Gaussian noise setting. First of all, replacing $(X,\beta,n,p,\alpha_k,C)$ in Theorem 4.2 of \citep{osher2014} by $(\Psi,\ga,l,m,l\triangle t,C\sqrt{l})$, we will directly get the Theorem \ref{thm:LB}, but under the assumption of Gaussian noise. To extend the result to sub-Gaussian noise setting, it only remains to modify the tail probability bound for $\Prob(\mathcal{A})+\Prob(\mathcal{B})+\Prob(\mathcal{C})$, where
	\begin{eqnarray*}
		\mathcal{A}&=&\{\epsilon:\|\Psi_S(\Psi_S^T \Psi_S)^{-1}\Psi_S^T \Psi\epsilon\|_2 > 2\sigma\sqrt{s\log l}\}, \\
		\mathcal{B}&=&\{\epsilon:\|(\Psi_S^T \Psi_S)^{-1}\Psi_S^T\Psi\epsilon\|_{\infty} > 2\sigma\sqrt{\frac{\log m}{C_{min}}}\},\\
		\mathcal{C}&=&\{\epsilon:\|\Psi_{S^c}^T (I-P_S)\Psi\epsilon\|_{\infty} > 2\sigma\sqrt{\mu_{\Psi}\log m}\}.
	\end{eqnarray*}

	In the proof of Theorem \ref{thm:LASSO} above, we have shown that $\Prob(\mathcal{B})+\Prob(\mathcal{C})\le\frac{2}{m}$. Now it remains to bound $\Prob(\mathcal{A})$.

	Let $W^T = \Psi_S(\Psi_S^T \Psi_S)^{-1}\Psi_S^T \Psi$, then $\mathrm{tr}(W^TW)= \mathrm{tr}(\Psi_S(\Psi_S^T \Psi_S)^{-1}\Psi_S^T) = \mathrm{tr}((\Psi_S^T \Psi_S)^{-1}\Psi_S^T\Psi_S) = s$. Note that
	\begin{eqnarray*}
		&&\left\{\epsilon:\|W\epsilon\|_{2} > 2\sigma\sqrt{\mathrm{tr}(W^TW)\log l}\right\}\\
		&\subseteq & \bigcup_j \{\epsilon:|W_j^T\epsilon| > 2\sigma\sqrt{\log l}\|W_j\|\},
	\end{eqnarray*}
	where $W_j\epsilon$ is a sub-gaussian variables with variance proxy $\sigma^2\|W_j\|^2$. Hence
	\begin{eqnarray*}
		&&\Prob(\{|W_j^T\epsilon| > 2\sigma\sqrt{\log l}\|W_j\|\})\\
		&\le& 2\exp\left( -\frac{4\sigma^2\log l\|W_j\|^2}{2\sigma^2\|W_j\|^2} \right)\\
		&=& 2l^{-2}
	\end{eqnarray*}
	Passing the union bound of index $j$ indicates that $\Prob(\mathcal{A}) \le \frac{2}{l}$.
\end{proof}

\end{document}